\documentclass[aps,a4paper,twocolumn,nofootinbib,superscriptaddress]{revtex4}

\usepackage[OT1]{fontenc}

\usepackage[english]{babel}

\usepackage[utf8]{inputenc}

\usepackage[per-mode=symbol,separate-uncertainty = true,multi-part-units=single]{siunitx}
\usepackage[version=3]{mhchem}
\usepackage{amssymb,amsfonts,mathrsfs}

\usepackage[amsmath,thmmarks]{ntheorem}

\usepackage{graphicx}

\usepackage{textcomp,marvosym}
\usepackage{fixltx2e}

\usepackage[linkcolor=black,colorlinks=true,citecolor=black,filecolor=black]{hyperref}
\usepackage{dsfont}

\def\ec{\emph{E.coli }}
\newcommand{\Drot}{D_{\mathrm{rot}}}
\DeclareSIUnit\Molar{\textsc{M}}

\begin{document}

\title{Statistical parameter inference of bacterial swimming strategies}

\author{Maximilian Seyrich}
\email[Email: ]{seyrich@tu-berlin.de}
\affiliation{Institut f\"ur Theoretische Physik, Technische Universit\"at Berlin, Hardenbergstrasse 36, 10623 Berlin, Germany} 

\author{Zahra Alirezaeizanjani}
\affiliation{Institut f\"ur Physik und Astronomie, Universit\"at Potsdam, Karl-Liebknecht-Strasse 24/25, D-14476, Germany}

\author{Carsten Beta}
\affiliation{Institut f\"ur Physik und Astronomie, Universit\"at Potsdam, Karl-Liebknecht-Strasse 24/25, D-14476, Germany}

\author{Holger Stark}
\email[Email: ]{holger.stark@tu-berlin.de}
\affiliation{Institut f\"ur Theoretische Physik, Technische Universit\"at Berlin, Hardenbergstrasse 36, 10623 Berlin, Germany}

\date{\today} 

\begin{abstract}
We provide a detailed stochastic description of the swimming motion of an \ec bacterium in two dimension, where we resolve tumble events in time. For this purpose, we set up two Langevin equations for the orientation angle and speed dynamics. Calculating moments, distribution and autocorrelation functions from both Langevin equations and matching them to the same quantities determined from data recorded in experiments, we infer the swimming parameters of \ec. They are the tumble rate $\lambda$, the tumble time $r^{-1}$, the swimming speed $v_0$, the strength of speed fluctuations $\sigma$, the relative height of speed jumps $\eta$, the thermal value for the rotational diffusion coefficient $D_0$, and the enhanced rotational diffusivity during tumbling $D_T$. Conditioning the observables on the swimming direction relative to the gradient of a chemoattractant, we infer the chemotaxis strategies of \ec. We confirm the classical strategy of a lower tumble rate for swimming up the gradient but also a smaller mean tumble angle (angle bias). The latter is realized by shorter tumbles as well as a slower diffusive reorientation. We also find that speed fluctuations are increased by about $30\%$ when swimming up the gradient compared to the reversed direction.
\end{abstract}

\maketitle

\section{Introduction}
One of the most prominent model swimmers in the field of biological microswimmers is the gut bacterium \ec
equipped with peritrichous flagella \cite{berg2008coli}. Its well known run-and-tumble swimming motion and chemotaxis strategy has been thoroughly studied \cite{adler1969chemoreceptors,berg1972chemotaxis,block1982impulse,schnitzer1993theory,
sourjik2012responding,turner2016visualizing}. Nowadays, modern imaging techniques allow for high-throughput recording of bacterial trajectories 
\cite{saragosti2011directional,masson2012noninvasive,cheong2015rapid,wu2006collective,
dufour2016direct,vater2014swimming,figueroa20183d,taute2015high}. 
The method of labeling flagella by fluorescent markers allows to unravel the diverse swimming mechanisms of 
microorganisms \cite{turner2000real,hintsche2017polar}. These refined techniques require an appropriate theoretical modeling of the bacterium's stochastic swimming path, including the dynamics of tumbling. They also require a rational and effcient method how to analyze the recorded data in experiments.
In this article we provide such a theoretical framework and illustrate it for the model bacterium \mbox{\ec.} Thereby, we also reveal some new and detailed insights into its chemotaxis strategy.


The \ec bacterium resides in the run phase, when all of its flagella form a bundle and rotate
counterclockwise. The bacterium swims along a straight line, only thermal rotational diffusion affects its persistence. When 
at least one of the flagella reverses its sense of rotation, it leaves the bundle and the bacterium is in the tumble phase, where it strongly reorients \cite{larsen1974change,macnab1974bacterial}. 
Typically, the tumble phase is much shorter than the run phase \cite{berg2008coli}. Therefore, in theoretical models tumbling is considered as
instantaneous and a single event is described by a tumble angle drawn from a distribution\ \cite{berg1993random,tailleur2008statistical,saragosti2011directional,pohl2017inferring}. However, a recent and instructive work by Saragosti \emph{et al.} showed that reorientation during tumbling can be modeled by enhanced rotational diffusion \cite{Saragosti2012}.

In order to analyze large amounts of data from recorded trajectories, specialized computer algorithms, called tumble recognizers, 
have widely been used to identify tumble events \cite{berg1972chemotaxis,saragosti2011directional, masson2012noninvasive}. In order to distinguish runs from tumbles, these automated tumble recognizers compare turning rate and speed to threshold parameters.  They are necessary to distinguish variations of speed and turning rate due to the ubiquitous noise from a real tumble event. The threshold parameters have to be chosen a-priori and adjusted until results from the automatized tumble recognition agree with a visual inspection of the trajectories. There is no general rule how to set these parameters
and indeed they vary quite substantially \cite{berg1972chemotaxis,saragosti2011directional}. 

In an earlier work \cite{pohl2017inferring}, we presented a parameter inference technique that allows to quantify the swimming behavior of bacteria without the need of setting parameters a priori.
Kramers-Moyal coefficients were calculated from a suitable stochastic model for the dynamics of the orientation angle and matched to the coefficients determined from experimental data. In particular, the stochastic model treated tumble events as instantaneous. This procedure provided the main characteristics of
\ec and the bacterium \emph{Pseudomonas putida}: tumble rate, distribution of tumble angles, and the thermal rotational diffusivity. For \ec it also confirmed an angle bias during chemotaxis reported earlier \cite{saragosti2011directional}: the mean tumble angle is larger when swimming against a chemical gradient compared to moving along it. Other parameter inference techniques use the framework of Bayesian inference \cite{masson2012noninvasive,rosser2013novel}. However, they pose a complex numerical challenge as one has to maximize a likelihood function that contains the data of all the recorded trajectories. 

\begin{figure}
\label{fig:Figure1}
    \includegraphics[width= 0.4\textwidth]{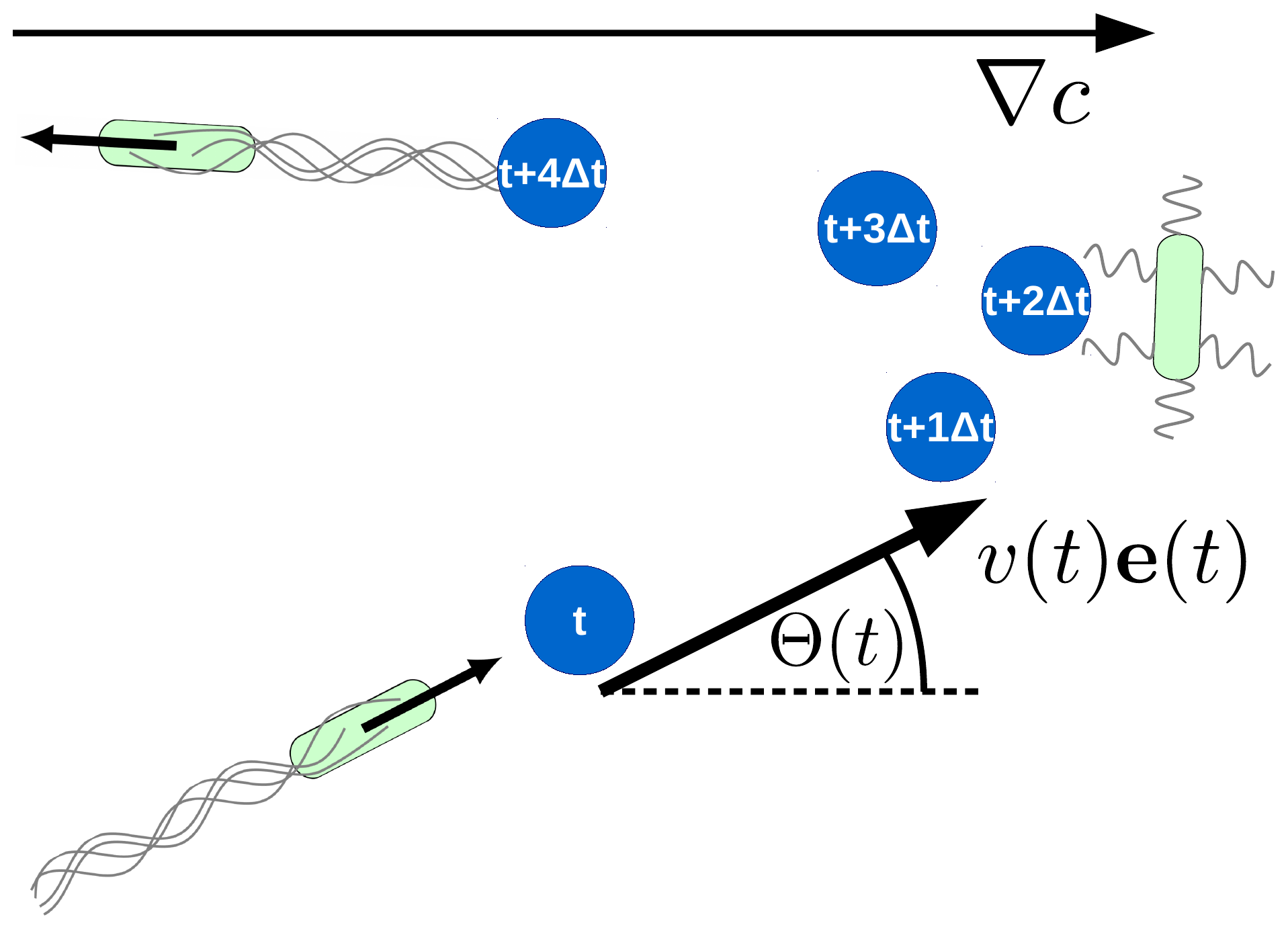}
    \caption{\ec with swimming velocity $\mathbf{v}(t)= v(t) \mathbf{e}(t) = v(t) [\cos\Theta(t) , \sin\Theta(t)]$. A tumble event occurs between the times $t+\Delta t$ and $t+3\Delta t$. A possible chemical gradient is indicated.}
\end{figure}

In this article we considerably extend our earlier work by resolving tumble events in time and by incorporating a stochastic process for the speed dynamics (see Fig.\ \ref{fig:Figure1}). The dynamics of the orientation angle is diffusive, where the rotational diffusivity switches via a telegraph process \cite{lindner2009brief} between its thermal (run phase) and enhanced value (tumble phase). The dynamics of the speed contains a  shot-noise process \cite{van1983relation,strefler2009dynamics}. It initiates a tumble event by decreasing the speed value, which then relaxes back to the swimming speed according to an Ornstein-Uhlenbeck process \cite{uhlenbeck1930theory}.
Calculating moments, distribution and autocorrelation functions for orientation angle as well as speed and matching them to the same quantities calculated from experimental data, we are able to infer the swimming parameters of \mbox{\ec.} 
Their values are in good agreement with the parameters determined using a tumble recognizer. Compared to the Bayesian framework, our method of parameter inference considerably lowers the efforts of the numerical optimization.

To explore the chemotaxis strategy of \ec, we condition \cite{graham1989stochastic,bandi2003functional} moments and autocorrelation functions on the swimming direction relative to the chemical gradient and infer the swimming parameters as a function of the orientation angle. Besides the well-known chemotaxis strategy (modulation of the tumble rate), we confirm the recently discovered angle bias \cite{saragosti2011directional}. We show that the increased angular persistence when swimming up the gradient is caused by both shorter tumbles as well as smaller rotational diffusivity. Moreover, for the same swimming direction we identify larger fluctuations in the speed value.

The article is organized as follows. In Sect.\ \ref{sec.method} we introduce the two Langevin equations of our stochastic model and calculate moments, distribution functions, and autocorrelation functions for speed and orientation angle. Section\ \ref{Sec.Materials} reviews details of the experiments. Section\ \ref{sec.results} first explains the inference method and then presents our results in a uniform buffer solution (control experiment) and in the gradient of a chemoattractant. We close with a summary and an outlook in Sect.\ \ref{sec.concl}.

\section{Model and Method}
\label{sec.method}

\subsection{Stochastic model for the random walk of \ec}

A typical trajectory of bacteria such as \ec is described by a run-and-tumble random walk. During the run phase the bacterium moves forward along a nearly straight line, only rotational thermal noise affects its persistence. During the tumble phase the bacterium's speed is reduced and it reorients strongly into a new direction. The angle between the orientations before and after the tumble event is the tumble angle $\beta$. We express the velocity of the bacterium in two dimensions as the product of speed $v(t)$ and unit vector $\mathbf{e}(t) =  (\cos\Theta ,\sin \Theta)$,
\begin{equation}
\dot{\mathbf{r}}(t) = v(t)\mathbf{e}(t) \, , 
\label{eq:r}
\end{equation}
where the orientation angle $\Theta$ is measured with respect to the $x$\ axis. 

We set up two overdamped Langevin equations for speed and orientation angle, which fully describe the bacterial motion,
\begin{align}
\dot{v}(t) &= r\left[v_0 - v(t)\right] + \xi_{\textrm{sp}}(t) + q(t), \label{eq:vt}\\
\dot{\Theta}(t) &= \sqrt{2\Drot(t)}~\xi_{\textrm{an}}(t) \, . \label{eq:theta}
\end{align}
We introduce both Langevin equations in more detail.

\begin{figure}
  \includegraphics[width=0.40\textwidth]{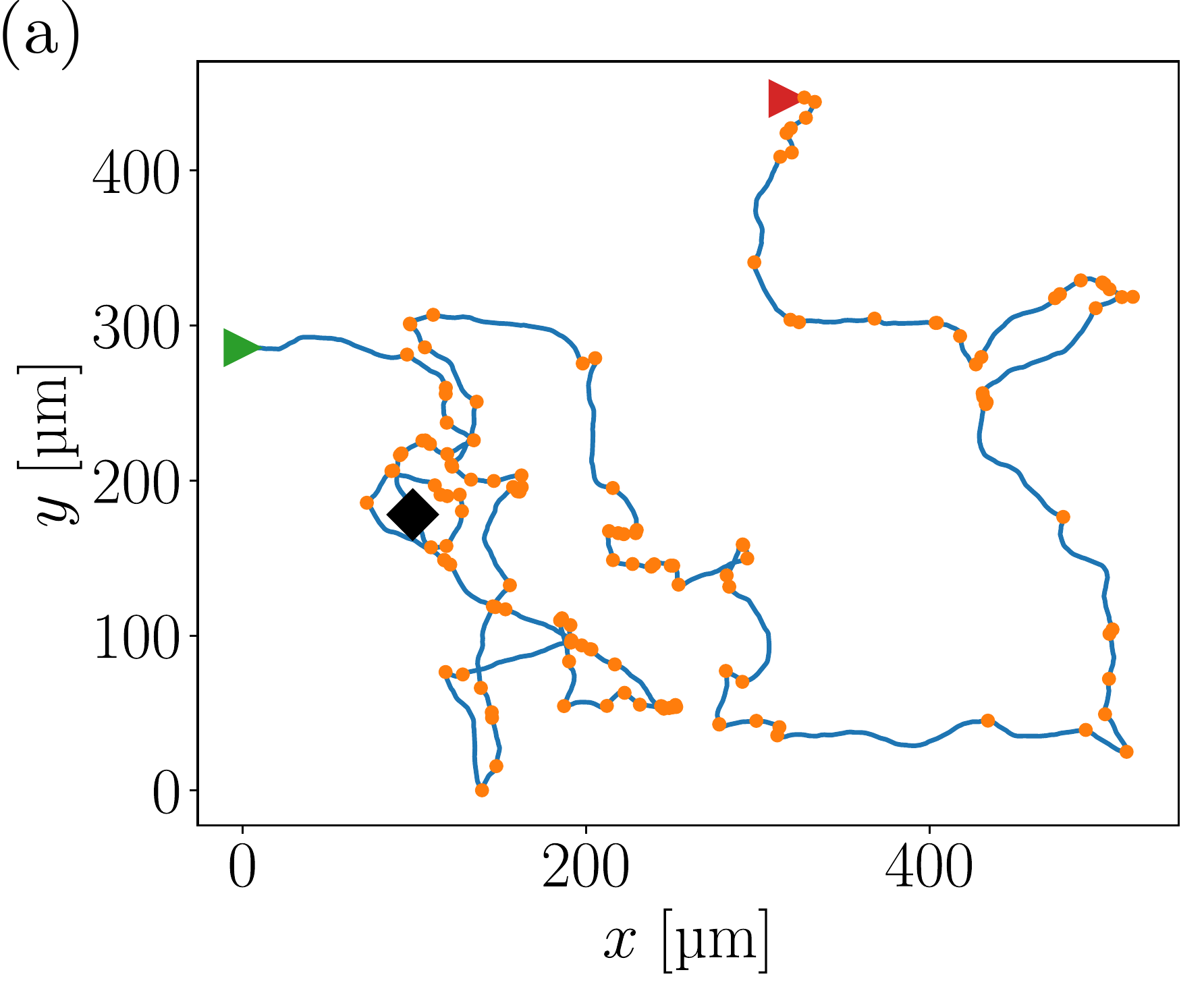}
 \includegraphics[width= 0.23\textwidth]{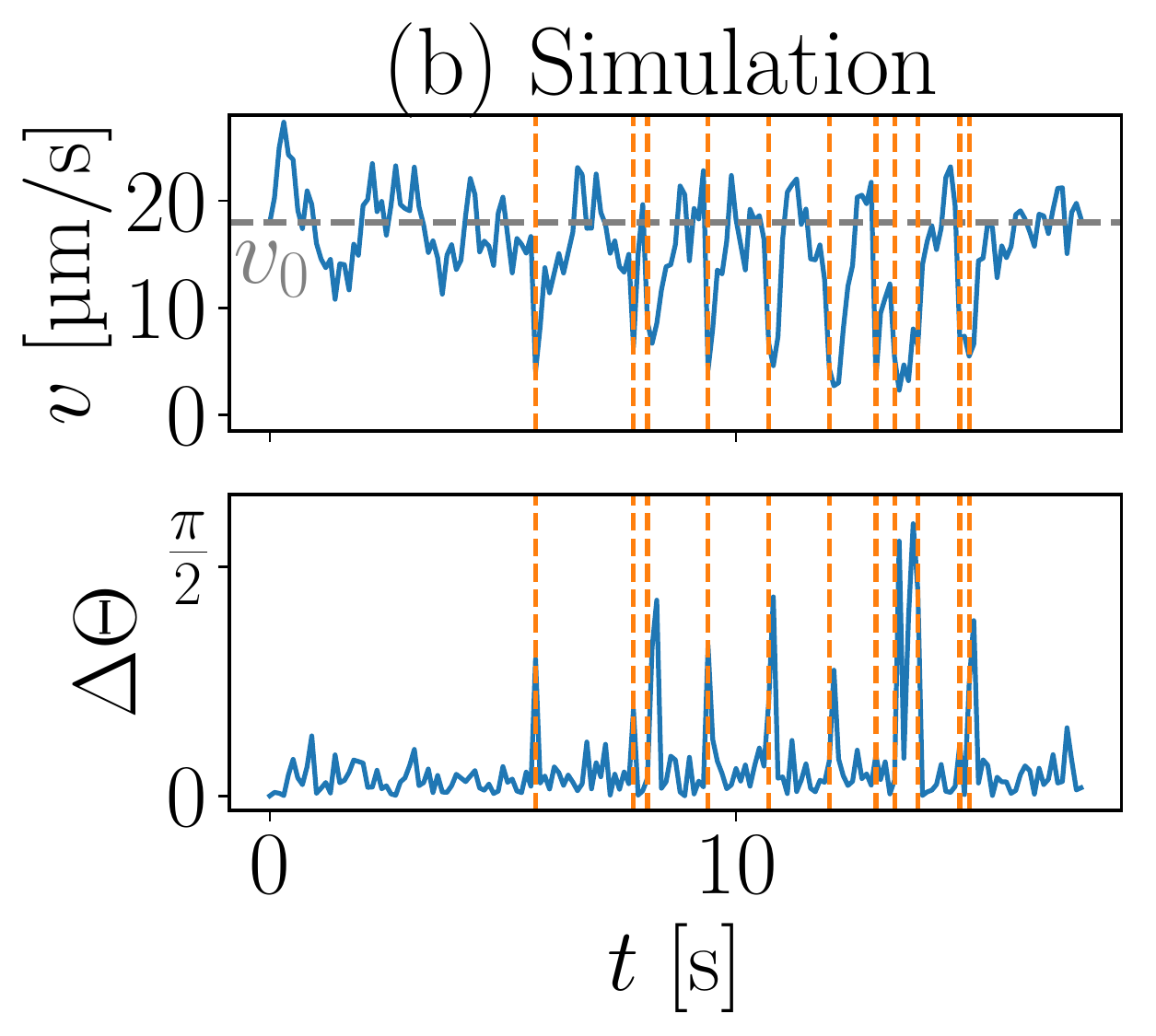}
 \hfill
  \includegraphics[width= 0.23\textwidth]{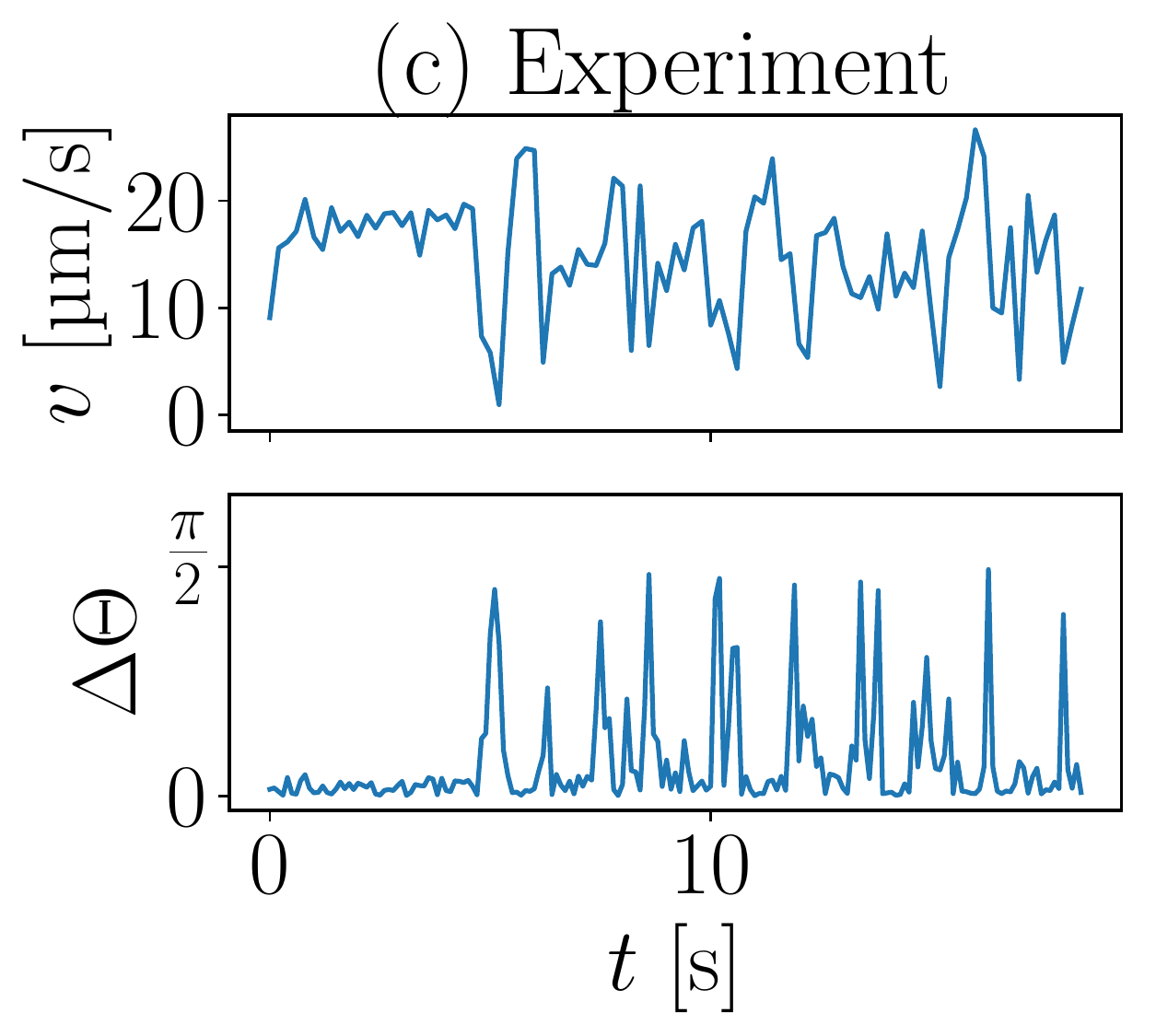}
   \caption{
   (a) Simulated run-and-tumble trajectory of a bacterium using the stochastic equations\ (\ref{eq:vt}) and (\ref{eq:theta}). It starts at the green and ends at the red triangle. 
   (b) Initial part (from green triangle to the black diamond)
   of the corresponding time series for speed $v(t)$ and angular displacement $\Delta\Theta (t)$ during time step $\Delta t = \SI{0.1}{s}$. Tumble initiations are marked in orange. (c) Experimental time series for $v(t)$ and $\Delta\Theta(t)$.
   }
  \label{fig:SimProcesses}
 \end{figure}

\textbf{(1)} 
The equation for speed $v(t)$ contains three terms, which are associated with drift, diffusion, and jumps.
We start with the last term,
\begin{equation}
q(t) = - \sum_{i=1}^{N^{\mathbin{{\lambda}}}}  \eta v(t) \delta(t-t_i) \, .
\label{eq:shot}
\end{equation}
It initiates each tumble event at time $t_i$ by a shot-noise process, while the occurrence of times $t_i$ follows a Poisson process with tumble rate $\lambda$. At the beginning of 
each tumble, the bacterial speed is reduced by the relative jump height $\eta$ to $(1-\eta)v_t$ and $N^{\lambda}$ is the actual number of tumble events. The first and second term represent a conventional
Ornstein-Uhlenbeck process. After a tumble event the speed relaxes with
relaxation rate $r$ towards the swimming speed $v_0$ of the run phase. Thus $r^{-1}$ is the mean duration of a tumble event, which we call tumble time in the following. The Gaussian white noise term is fully determined by $\langle  \xi_{\textrm{sp}} \rangle = 0$ and $\langle \xi_{\textrm{sp}}(s) \xi_{\textrm{sp}} (t) \rangle = \sigma^2 \delta(t-s)$, where we introduce the white noise strength $\sigma$. It describes the ubiquitous noise due to internal noise of the swimming mechanism, variations between individual bacteria, and measurement errors. Note that the actual tumble time of a bacterium is exponentially distributed. In our model the white noise term 
also induces stochastic fluctuations in the duration of the tumble events as visible in Fig.\ \ref{fig:SimProcesses}(b).
Altogether, the stochastic speed process is determined by five parameters: $\{ \lambda, r , v_0,  \sigma, \eta\} $.

\textbf{(2)} The stochastic equation for the orientation angle $\Theta$ is fully described by rotational diffusion, where the white noise process is defined by $\langle \xi_{\textrm{an}} \rangle = 0$ and $\langle \xi_{\textrm{an}}(s) \xi_{\textrm{an}}(t) \rangle = \delta(t-s)$. Following Ref.\ \cite{saragosti2012modeling}, we model tumbles as a random walk on a unit sphere with enhanced rotational diffusion. Thus, the rotational diffusion coefficient $D_{\textrm{rot}}(t)$ is no longer a constant but alternates between two values: the thermal rotational diffusion coefficient $D_0$ during run phases and an enhanced value $D_T$ during tumble phases. 
We describe each transition between the two states by a Poisson process and thus obtain a telegraph process. The transition rate from the run to the tumble phase is the tumble rate $\lambda$, whereas the transition rate in the opposite direction is the speed relaxation rate $r$ or the inverse tumble time. A full definition and basic properties of the telegraph process are given in the appendix \ref{AppendixAngle} or can be found in \cite{lindner2009brief}.

To link the telegraph process to the shot-noise process for the speed value in Eq.\ (\ref{eq:shot}), the diffusion coefficient switches at the same times $t_i$ from the thermal ($D_0$) to the enhanced ($D_T$) value. Note, while the speed process allows a second tumble although the first one is not finished yet, this is not possible in the telegraph process for rotational diffusion. However, for bacteria like \ec the time between tumble events is typically one order of magnitude larger than the tumble time $r^{-1}$. This makes these double events very rare and tumble events in both speed and angular processes coincide. All in all, we have four parameters governing the stochastic process for the orientation angle: $\{\lambda, r, D_0, D_T\}$.

Figure \ref{fig:SimProcesses} shows a typical simulated trajectory (a) and the corresponding time series for speed and angular displacement $\Delta \Theta$
during time step $\Delta t = \SI{0.1}{s}$ (b). It has to be compared to the experimental time series of both quantities in (c). Note 
that $\tfrac{\Delta\Theta}{\Delta t}$ represents the turning rate of the bacterium. In the following we will always work with the angular displacement $\Delta \Theta$.

\subsection{Basics of the inference method}

In this section we state moments, stationary distributions, and time autocorrelation functions for the stochastic processes of
speed and orientation angle in Eqs.\ (\ref{eq:vt}) and (\ref{eq:theta}). They depend on the swimming parameters introduced above. 
Matching the theoretical expressions of these quantities to the values determined by averaging over all individual tracks of the 
experiments, we are able to infer the mean swimming parameters of an \ec population. We refer to appendix \ref{AppendixMoments} 
for details of the derivations and only state the final expressions in the following. 

\begin{figure}

 \includegraphics[width=0.49\textwidth]{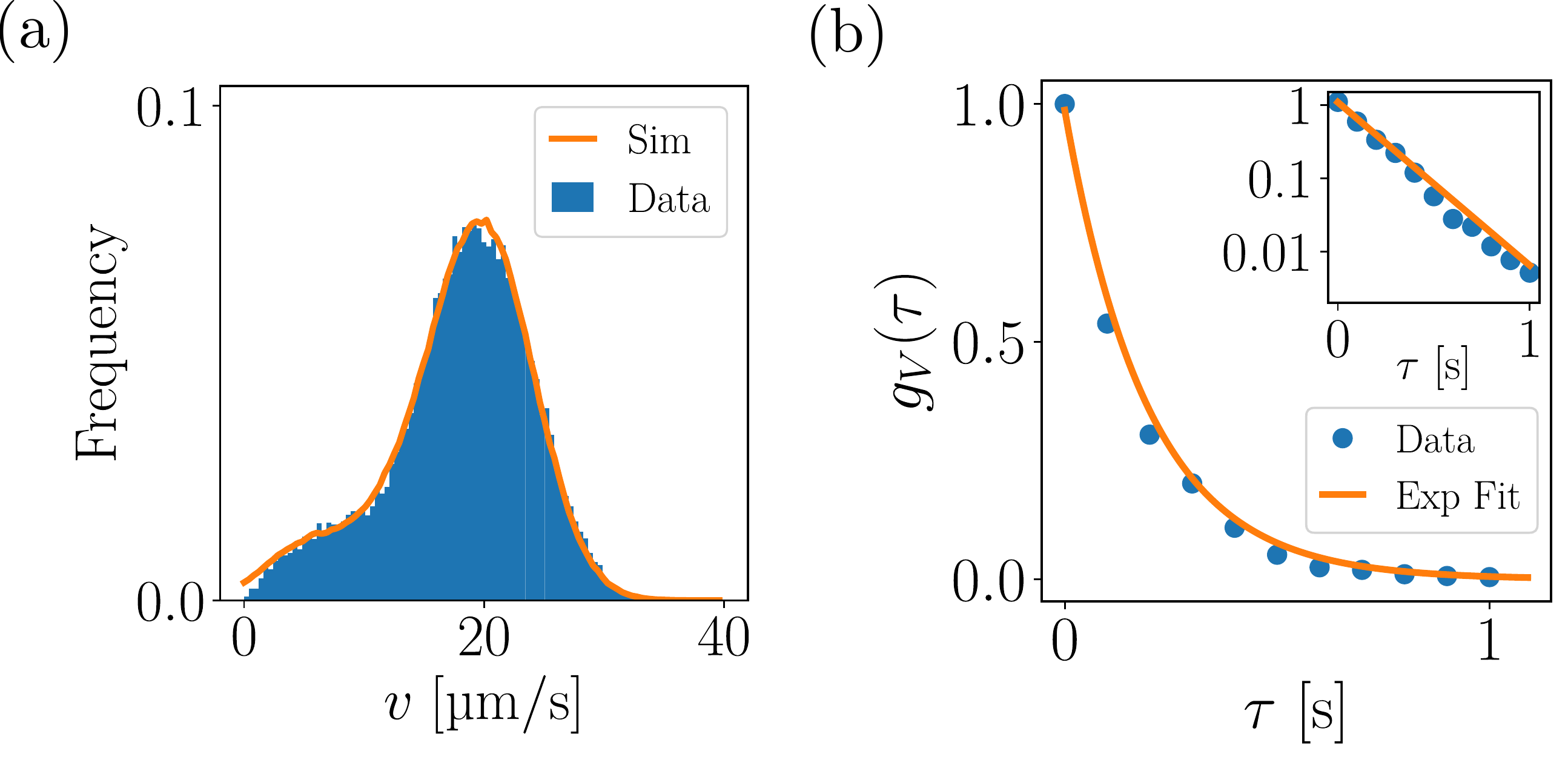}
  \caption{(a) Histogram showing the distribution of speed values for a dataset recorded for \ec 
  in a control experiment moving in a buffer medium without any chemical gradient. The orange line shows the distribution from the simulated process using the inferred parameters. (b) Corresponding speed autocorrelation function $g_V(\tau)$ of the same dataset. The orange line shows an exponential fit with   relaxation rate $\alpha_V =  \SI{5.1 \pm 0.2}{s^{-1}} $.
  Inset: Semi-logarithmic plot of $g_V(\tau)$.
} 
  \label{fig:ExpHistAutocorrSpeed}
\end{figure}

\subsubsection{Speed}
The moments $m^V_n  =  \langle v(t) ^ n\rangle$ of Eq.\ (\ref{eq:vt}), where the average is taken over all times $t$ and all tracks in the long-time limit, can be calculated 
as a function of the reduced parameter set $\left( \lambda / r , \eta, v_0, \sigma^2 / r \right)$. For the first moment, the mean speed, we obtain
\begin{equation}
\label{eq:SpeedMoment1}
m^V_1 \left( \frac{\lambda}{ r}, \eta, v_0, \frac{\sigma^2}{r} \right) = \frac{v_0}{1+\eta\lambda/r}~.
\end{equation}
The mean speed is smaller than the swimming speed $v_0$ since during the tumble phase speed is reduced by a factor $\eta$. More generally, a recursive formula for the $n$th moment is given by
\begin{equation}
\label{eq:SpeedMomentN}
m^V_n \left( \frac{\lambda}{ r}, \eta, v_0, \frac{\sigma^2}{r} \right) 
= \frac{v_0 ~m_{n-1}^V ~+~ \frac{1}{2} \left(n-1\right) \frac{\sigma^2}{r} ~m_{n-2}^V  } 
       {1 + \frac{\lambda}{n r} - \frac{\lambda}{nr} \left( 1-\eta \right) ^n } ~,
\end{equation}
where the zeroth moment is $m_0 =1$ due to normalization. We now have access to all the speed moments. 
As an example, Fig.\ \ref{fig:ExpHistAutocorrSpeed}(a) shows a histogram for the distribution of speed values recorded in an experiment, from which the speed moments can be calculated. The orange line represents the distribution obtained from numerically solving the speed equation (\ref{eq:vt}) using the actual 
parameters inferred from this experiment. The two distributions nicely agree, which is an a-posteriori verification of our Langevin equation.

From the moments we can only infer the ratios $\lambda / r$ and $\sigma^2/r$. In order to determine the full set of parameters of Eq.\ (\ref{eq:vt}), we also use the speed autocorrelation function for our model. It has an exponential form with relaxation rate $r+\eta\lambda$,
\begin{equation}
\label{eq:SpeedAutocorr}
g_V(\tau) = \langle [v(t+\tau) - m^V_1] [v(t) - m^V_1] \rangle = {\Delta ^2 v }{e^{-(r+\eta\lambda)\tau}} ~,
\end{equation}
where we have introduced the variance $\Delta^2 v = \langle (v - m^V_1)^2 \rangle $.
Figure\ \ref{fig:ExpHistAutocorrSpeed}(b) shows the autocorrelation function for the experimental data of \ec. Indeed, the curve is well-fitted by an exponential over two decades up to $\tau \simeq 1s$, which is around half the mean track length. This agreement supports the validity of our stochastic description of the speed process in Eq.\ (\ref{eq:vt}).

\subsubsection{Angle}

\begin{figure}
\includegraphics[width=0.49\textwidth]{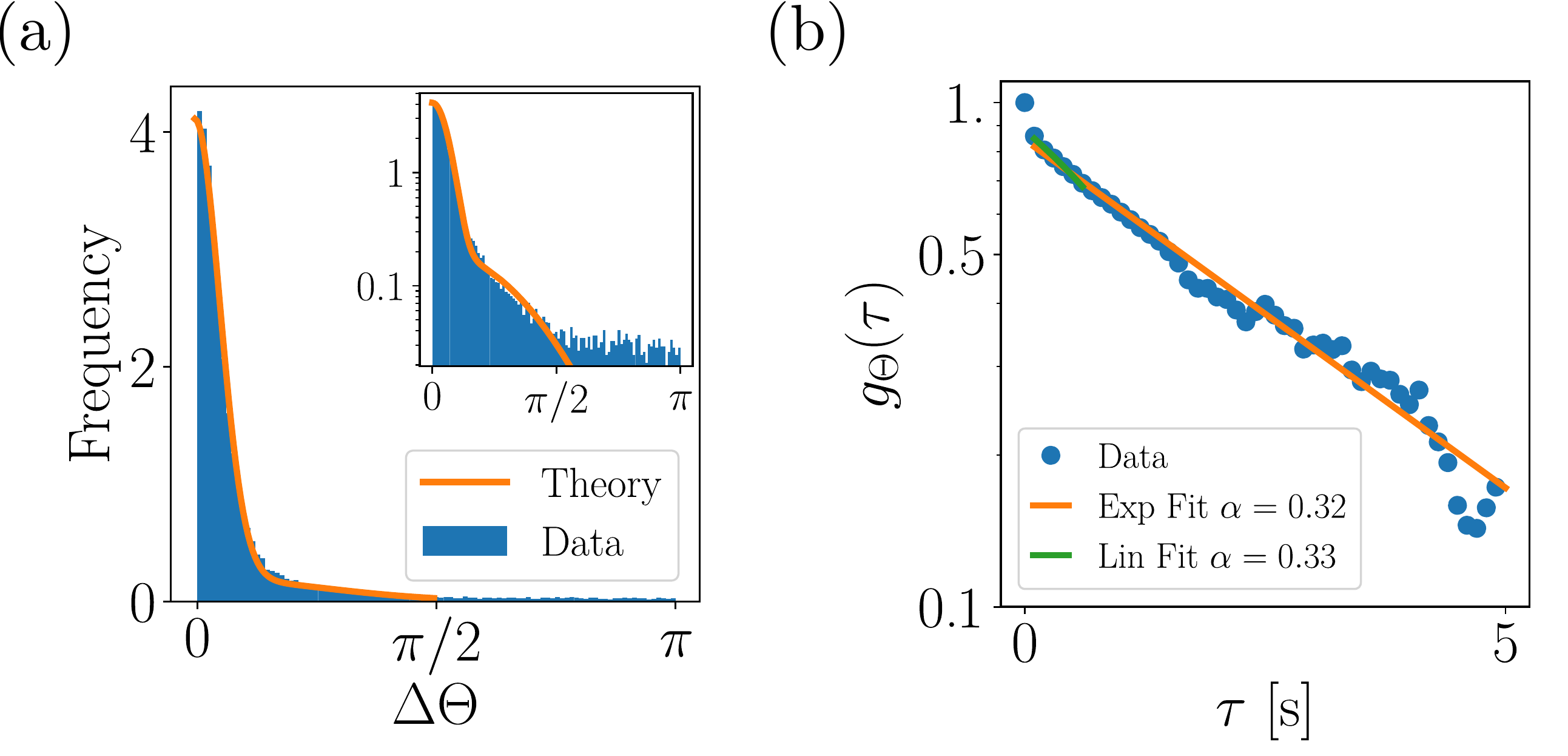}
  \caption{(a) Histogram showing the distribution of angular displacements $\Delta \Theta$ in time step $\Delta t$ for the same data set as in Fig.\ \ref{fig:ExpHistAutocorrSpeed}.
The orange line shows the distribution $p(| \Delta \Theta |)$ from Eq.\ (\ref{eq:pdfTurnrate}) using the inferred parameters. Inset: Semi-logarithmic plot of the distribution. (b) Semi-logarithmic plot of the corresponding directional autocorrelation function $g_{\Theta}(\tau)$.
Green line: linear fit with negative slope $\alpha_{\Theta} =\SI{0.33}{s^{-1}}$;
orange line: exponential fit with relaxation rate $\alpha_{\Theta} = \SI{0.32}{s^{-1}}$.}
 \label{fig:HistogramAutocorrAngle}
\end{figure}

Here, we work directly with the steady-state probability distribution $p(|\Delta \Theta|)$ for the absolute angular displacement $|\Delta \Theta|$ during a finite time step $\Delta t$. 
We determine $p(|\Delta \Theta|)$
from Eq.\ (\ref{eq:theta}) for the orientation angle
as a function of the reduced parameter set $(\lambda / r, D_0, D_T)$. 
In the long-time limit the probability distribution $p(|\Delta \Theta|)$ becomes stationary and is given by
\begin{equation}
\label{eq:pdfTurnrate}
 p( | \Delta \Theta | ) = \frac{r}{\lambda + r} \mathcal{N}(0,\, \sqrt{2D_0 \Delta t}) 
 + \frac{\lambda}{\lambda + r} \mathcal{N}(0,\,\sqrt{2D_T \Delta t})
\end{equation}
where $\mathcal{N}(0,\sigma)$ denotes the normal distribution with zero mean and standard deviation $\sigma$. 
For our pa\-ra\-me\-ter inference we use the same time step $\Delta t = 0.1 \textrm{s}^{-1}$ as in Ref. \cite{saragosti2012modeling}.

Figure\ \ref{fig:HistogramAutocorrAngle}(a) presents a histogram for all angular displacements
in time step $\Delta t$ recorded in the experiment. It shows a deviation from the theoretical distribution of Eq.\ (\ref{eq:pdfTurnrate}) in the tail at angles larger than $\pi/2$, which is visible only in the semi-logarithmic plot. Note that the region $|\beta|> \pi/2$ only represents roughly $3\%$ of all angular displacements. There are two possible reasons for this deviation: First, we record angular displacements $\Theta=\pi + \epsilon$ as a displacement $-(\pi - \epsilon)$ since we cannot distinguish between tumbles to the right and left during one time step. Second, it is also possible that the diffusion model for tumbling does not apply for such large angles.

For completeness we also give the $n$th moment of the absolute angular displacement,
$m^{\Delta \Theta}_n = \left \langle | \Delta \Theta |^n \right\rangle$. It follows directly from the probability distribution of Eq.\ (\ref{eq:pdfTurnrate}):

\begin{align} 
\label{eq:TurnrateMomentN}
\begin{aligned}
m^{\Delta \Theta}_n (\lambda,r,D_0,D_T) = &\left( \frac{(2D_0 \Delta t)^{\frac{n}{2}}}{1+\lambda /r} + \frac{(2D_T \Delta t)^{\frac{n}{2}}}{1+r/\lambda} \right) (n-1)!! \\
&\cdot\begin{cases} \sqrt{\frac{2}{\pi}} ~~~&\textrm{if n is odd}\\ 1 ~~~&\textrm{if n is even} \end{cases} ~,
\end{aligned}
\end{align}
where $n!!$ denotes the double factorial.

Similar to the speed process, we can only infer the ratio $\lambda / r$ from fits to the probability distribution $p(|\Delta \Theta |)$ of Eq.\ (\ref{eq:pdfTurnrate}).
In order to determine the full set of parameters of Eq.\ (\ref{eq:theta}),
we use again the autocorrelation function of our model, now for the swimming direction $\bold{e}(t)$. Numerical investigations of our model (see appendix \ref{NumericalVerifications}) suggest that it has a simple exponential form with relaxation rate $\alpha_{\Theta}$ for parameters relevant to the experiments:
\begin{equation}
\label{eq:AngleAutocorr}
g_{\Theta}(\tau) = \left\langle \bold{e}(t+\tau)\cdot \bold{e}(t)\right\rangle \propto e^{-\alpha_{\Theta} \tau}
\end{equation}
Analytically, we are not able to calculate this exponential form. However, in the time interval 
$ (\lambda+r)^{-1}  <  \tau <  \langle D_{\rm{rot}} \rangle^{-1} $ relevant to the experiments,
we can derive the linear approximation
\begin{equation}
\label{eq:LinearApproxAngleAutocorr}
   g_\Theta(\tau) \approx  1 -  \alpha_{\Theta} \tau    
         \approx 1- \left( \langle D_{\rm{rot}} \rangle  - \frac{\Delta ^2 D_{\rm{rot}}}{\lambda+r} \right) \tau 
\end{equation}
and thereby obtain an expression for the relaxation rate $\alpha_{\Theta}$. Here we have introduced the 
respective mean $\langle D_{\rm{rot}} \rangle$ and variance $\Delta^2 D_{\rm{rot}}$ of the telegraph process $D_{\rm{rot}}(t)$,
\begin{eqnarray}
\langle D_{\rm{rot}}\rangle & = & \frac{D_0}{1+\lambda/r} + \frac{D_T}{1+r/\lambda}  \nonumber \\
\Delta^2 D_{\rm{rot}} & = & \langle (D_{\rm{rot}} - \langle D_{\rm{rot}}\rangle)^2 \rangle =  \frac{(D_0-D_T)^2 \lambda/r }{(1+ \lambda/r)^2}  \, .
\end{eqnarray}

Figure\ \ref{fig:HistogramAutocorrAngle}(b) shows the directional autocorrelation function for the experimental data of \ec moving in a uniform buffer medium.
Indeed, the curve is well-fitted by an exponential up to $\tau \simeq \SI{5}{s}$ excluding the first point. This agreement supports the validity of our stochastic description of the angle process in Eq.\ (\ref{eq:theta}). The deviation in the experimental data for the first point is caused by the offset for angular displacements larger than $\pi/2$, where the experimental distribution function in Fig.\ \ref{fig:HistogramAutocorrAngle}(a) deviates from theory. For two and more time steps the influence of this offset becomes smaller and smaller.

\section{Experimental materials and methods}
\label{Sec.Materials}

\subsection{Cell culture}
{\ec}AW405 strain was cultured overnight in liquid Tryptone Broth (TB) (\SI{10}{\gram\per\litre} Difco Bacto\textsuperscript{TM}-Tryptone and \SI{5}{\gram\per\litre} \ce{NaCl}) at \SI{37}{\celsius} on a rotary shaker at \SI{300} rpm. The cell suspension was diluted 1:100 into fresh TB, and grown to mid-exponential phase (OD\textsubscript{600} = 0.5).
Then the bacterial suspension was washed and resuspended in motility buffer (\SI{11.2}{\gram\per\litre} \ce{K2HPO4}, \SI{4.8}{\gram\per\litre} \ce{KH2PO4}, \SI{3.93}{\gram\per\litre} \ce{NaCl}, \SI{0.029}{\gram\per\litre} EDTA and \SI{0.5}{\gram\per\litre} glucose; pH 7.0). Afterward, the cell suspension was divided into two fractions. One was centrifuged and resuspended in the same motility buffer, and the other was centrifuged and resuspended in motility buffer supplemented with the chemoattractant $\alpha$-methyl-aspartate (Sigma-Aldrich, USA) in a final concentration of \SI{0.5}{\milli\Molar}. In both cases, the final OD\textsubscript{600} of the cell suspensions was 0.07 before filling them into chemotaxis chambers.

\subsection{Chemotaxis assay}
In this study, a µ-Slide Chemotaxis 3D (ibidi, Martinsried, Germany) was used in order to maintain a stable linear gradient of the chemoattractant $\alpha$-methyl-aspartate. This chemotaxis chamber consists of two large reservoirs connected to a central observation area. For the chemotaxis assay, the cell suspension with chemoattractant was filled into the reservoir on the right hand side and the chemoattractant-free cell suspension into the reservoir on the left hand side. The central observation area was filled with motility buffer (see appendix \ref{AppendixIbidiChamber}). A stable linear chemoattractant gradient is generated by diffusion in the observation area and maintained for several hours \cite{zengel2011mu}. For the control assay, both reservoirs were filled with chemoattractant-free cell suspension. In this case, a homogeneous environment without any gradient was established in the observation area.

\subsection{Cell imaging and tracking}
An IX71 inverted microscope with a 20× UPLFLN-PH objective (both Olympus, Germany) in phase contrast mode was used for imaging cell trajectories. Five image sequences were taken with \SI{10}{\minute} intervals between them using a Orca Flash 4.0 CMOS camera (Hamamatsu Photonics, Japan). For each sequence, the images were acquired at 20 frames per second for \SI{30}{\second}. The field of the view was placed in the center of the gradient region at \SI{30}{\micro\metre} above the bottom of the chamber (total height in the observation area was \SI{70}{\micro\metre}).

A custom Matlab program based on the Image Processing Toolbox (version R2015a, The MathWorks, USA) was used to process the image sequences automatically. For each image sequence, a background image was calculated by pixel-wise time average projection. It was subtracted from each frame to eliminate non-motile objects and shading effects. The built‐in Matlab function \textit{imerode} was then applied for morphological erosion (with a disk of radius \SI{0.6}{\micro\metre}) to reduce the background noise. The putative bacterial cells are distinguished from background using the maximum entropy thresholding algorithm by Kapur et al. \cite{Kapur1985}. The threshold was calculated for each image in the sequence separately. The median of all threshold values was used to segment the whole sequence. The binary images were further processed with the morphological operations, \textit{imopen} and \textit{imclose} (with a disk of radius \SI{0.3}{\micro\metre}) to eliminate any noise caused by segmentation. The built‐in function \textit{bwconncomp} was used to find all connected objects in the binary images. Size and centroid of the objects were determined using the \textit{regionprops} function. Afterwards, particles with an area between \SIrange{1}{15.6}{\micro\metre^2} were considered as single bacterial cell. Finally, trajectories were obtained employing the tracking algorithms by Crocker and Grier \cite{crocker1996methods}.

To avoid tracking artifacts caused by tumble events when cells enter and leave the focal plane, the first and last \SI{0.5}{\second} of each track were removed. Highly curved tracks as well as tracks with a total displacement $<$ \SI{10}{\micro\meter} were eliminated, as they most likely result from damaged flagella. The minimal track length is \SI{0.5}{\second} and the maximal length is \SI{19.35}{\second}. The control data set consists of $769$ tracks with a total length of $\SI{1629}{s}$. The gradient data set consists of $3498$ tracks with a total length of $\SI{7206}{s}$.

\subsection{Heuristic run-tumble analysis}
\label{sec.heuristic} 
The trajectories were smoothed using a second-order Savitztky–Golay filter with a window size of 5 data points corresponding to 250 ms \cite{savitzky1964smoothing}. Instantaneous speed $v=\tfrac{\Delta s}{\Delta t}$, direction of propagation $\theta$, and turning rate $\omega = \tfrac{\Delta \theta}{\Delta t}$ were evaluated on the smoothed tracks. The tumble events were detected as described previously \cite{masson2012noninvasive,theves2013bacterial,pohl2017inferring}. Briefly, in the time series of speed and turning rate, local minima and local maxima were detected, respectively, to identify tumble events. Four parameters, two for the speed and two for the turning rate, were adjusted such that the recognition of tumble events was correct as checked by visual examination (threshold parameters $\alpha = 3$ and $\beta = 6.5$ and tumble duration parameters 0.55$\times\Delta v$ and 0.65$\times\Delta \omega$, see the Supporting Information S5 in Ref \cite{pohl2017inferring}).

\section{Results}
\label{sec.results}

We are now equipped to infer the swimming parameters from experimental data for different experimental settings. We first 
illustrate the inference method by applying it to a control experiment, where \ec swims in a homogeneous buffer solution. 
We validate the inference method by comparing the inferred parameters to their values determined by a heuristic tumble recognizer. Then we demonstrate that our method also reveals the chemotaxis strategy of \ec when moving in a chemical gradient. In particular, we apply it to data, which was recorded in a linear gradient of $\alpha$-methyl-aspartate.

\subsection{Inferring the swimming parameters for \ec\\ in a uniform environment}
\label{Controlexperiment}
Figures\ (\ref{fig:ExpHistAutocorrSpeed}) and (\ref{fig:HistogramAutocorrAngle}) show distributions and autocorrelation functions for speed and angular displacements recorded for \ec when swimming
in a homogeneous buffer without any chemical gradient. 
Note that speed and angle inference are performed separately from each other but they
are linked by the tumble rate $\lambda$ and the inverse tumble time $r$.

\subparagraph{Speed inference:}
From the histogram of the recorded speed values in Fig.\ \ref{fig:ExpHistAutocorrSpeed}(a) we 
determine the moments of the experimental speed data: $m_n^{v,\mathrm{exp}} :=  
N^{-1} \sum_{i=0}^N T_i^{-1} \sum_{t=0}^{T_i} \left[v_i(t)\right]^n $. 
The sums are taken over all tracks $i= 0,\dots, N$ and all times $t$,
where $T_i$ is the length of track $i$. Figure\ \ref{fig:ExpHistAutocorrSpeed}(b) shows the  exponential fit to the speed auto-correlation function, which yields the experimental relaxation rate $\alpha_V =  \SI{-5.1 \pm 0.2}{s^{-1}}$. Note that the error estimate and all the following ones are obtained by the method of bootstrapping (see appendix \ref{AppendixBootstrapping} for more details). We match the first eight speed moments and the relaxation rate to the their theoretical expressions of Eqs.\ (\ref{eq:SpeedMoment1})-(\ref{eq:SpeedAutocorr}) and obtain 9 non-linear equations for the speed swimming parameters. We solve these equations numerically using a simplex-downhill optimization algorithm from the python package \textit{scipy}.

\subparagraph{Angle inference:} 
Independently, we match the theoretical distribution function for the angular displacement [given in Eq.\ (\ref{eq:pdfTurnrate})]
to the experimentally recorded histogram in Fig.\ \ref{fig:HistogramAutocorrAngle}(a) and thereby extract the parameters $D_0$, $D_T$, and $\lambda/r$.
We perform the fit up to ${\Delta\Theta} = \pi /2 $ to avoid the offset for angular displacements larger than $\pi/2$. 
Last, by matching the experimental relaxation rate $\alpha_{\Theta}$ of the directional autocorrelation function to the theoretical expression of Eq.\ (\ref{eq:LinearApproxAngleAutocorr}), we obtain
the full set of parameters [see also Eqs.\ (\ref{eq:lambdaAngle}) and (\ref{eq:rAngle}) in appendix \ref{AppendixAngle}].
Figure \ref{fig:HistogramAutocorrAngle}(b) shows the linear fit with relaxation rate
$\alpha_{\Theta} = \SI{0.33}{s^{-1}}\pm 0.02$ (green line) and the exponential fit with  
rate $\alpha_{\Theta} = \SI{0.32}{s^{-1}} \pm 0.01$ (orange line) in a semi-logarithmic plot.

\subparagraph{Inferred Parameters:}
Table \ref{tab:Results1} gives an overview of the inferred swimming parameters for
the two stochastic processes for speed and angle. The two inferred
tumble rates $\lambda$ are very close together and the inverse tumble times $r$ agree within the error bars. 
Our results are in good agreement with tumble rate $\lambda = \SI{0.84}{\second^{-1}}$ and swimming velocity $v_0 = \SI{20.7}{\micro m s^{-1}}$ determined with a heuristic tumble recognizer (see Sect.\ \ref{sec.heuristic} and Ref.\ \cite{pohl2017inferring}). This validates our inference method. Moreover, our findings are in good agreement with previously measured tumble rates \cite{berg1972chemotaxis, masson2012noninvasive,pohl2017inferring} and swimming speeds \cite{son2016speed}. The inferred value for the thermal rotational {diffusivity
$D_0$ agrees with previously reported values in the literature, which range from $\SI{0.06}{s^{-1}}$ \cite{berg1993random,pohl2017inferring} to $\SI{0.18}{s^{-1}}$ \cite{Celani1391}.

We use the enhanced rotational diffusion coefficient $D_T = \SI{2.31}{s^{-1}} $ and the inverse tumble time $r = \SI{3.81}{s^{-1}}$ of the angle stochastic process to determine the distribution function of absolute tumble angles, $P(|\beta|)$, by recording the angular displacement for exponentially distributed tumble times with mean $r^{-1}$. The corresponding three-dimensional distribution function is obtained by multiplying the two-dimensional quantity with $\sin\beta$ from the solid angle element. The resulting distribution is shown in orange in Fig.\ \ref{fig:TumbleAngleDistr} for $|\beta| < \pi$. 
It has a maximum at $\beta_{\mathrm{max}} = 0.78 = \ang{45}$ and the mean tumble angle is $\langle |\beta| \rangle = 1.06 = \ang{61}$, which are remarkably close to the values $\beta_{\mathrm{max}} = \ang{45}$ and $\langle |\beta| \rangle = \ang{62}$ from Ref.\ \cite{berg1972chemotaxis}. The shape of the distribution function is similar to the one obtained with the heuristic tumble recognizer (blue bars). Also, the maximum values are very close. While the main characteristics of the two curves agree well, the heuristic tumble recognizer determines more tumbles for angles close to $\pi$. As a result, it finds a larger mean tumble angle $\langle |\beta|\rangle = 1.43= \ang{82}$. This might be explained as follows. Some tumbles occur only in one time interval, where one cannot distinguish between a leftward tumble angle $\tilde{\beta}$ and a rightward tumble $|\tilde\beta-2\pi|$. Thus, the heuristic tumble recognizer chooses always the smaller angle and, therefore, the distribution of tumble angles close to $\pi$ is enhanced. In contrast, our inference for the angle process only uses angular displacements up to $\pi/2$ in Fig.\ \ref{fig:HistogramAutocorrAngle}. Thus, it gives a more correct account of the distribution. 

\begin{table}[]
\centering
\begin{tabular}{lllll}
\toprule
\multicolumn{2}{l}{\centering \bf{Speed}}                       &\quad  & \centering \bf{Angle} &                            \\ \colrule
$\lambda$                   & \SI{0.83\pm 0.04}{s^{-1}}         & $~~~~~~~~$\quad & $\lambda$             & \SI{0.84\pm 0.02}{s^{-1}}  \\
$r$                         & \SI{4.41\pm 0.3}{s^{-1}}          &  & $r$                   & \SI{3.81\pm 0.3}{s^{-1}}   \\
$v_0$                       & \SI{20.8\pm 0.2 }{\micro ms^{-1}} &  & $D_0$                 & \SI{0.09\pm 0.002}{s^{-1}} \\
$\sqrt{\frac{\sigma^2}{r}}~$\quad  & \SI{5.11\pm 0.07}{\micro ms^{-1}} &  & $D_T$                 & \SI{2.31\pm 0.12}{s^{-1}}  \\ 
$\eta$                      & \SI{0.85 \pm 0.01}                &  &                       &                           \\
\botrule
\end{tabular}
\caption{Inferred parameters for the stochastic processes of speed and angle for \ec moving in a
buffer medium without a chemical gradient (control experiment).}
\label{tab:Results1}
\end{table}

\begin{figure}
    \includegraphics[width= 0.4\textwidth]{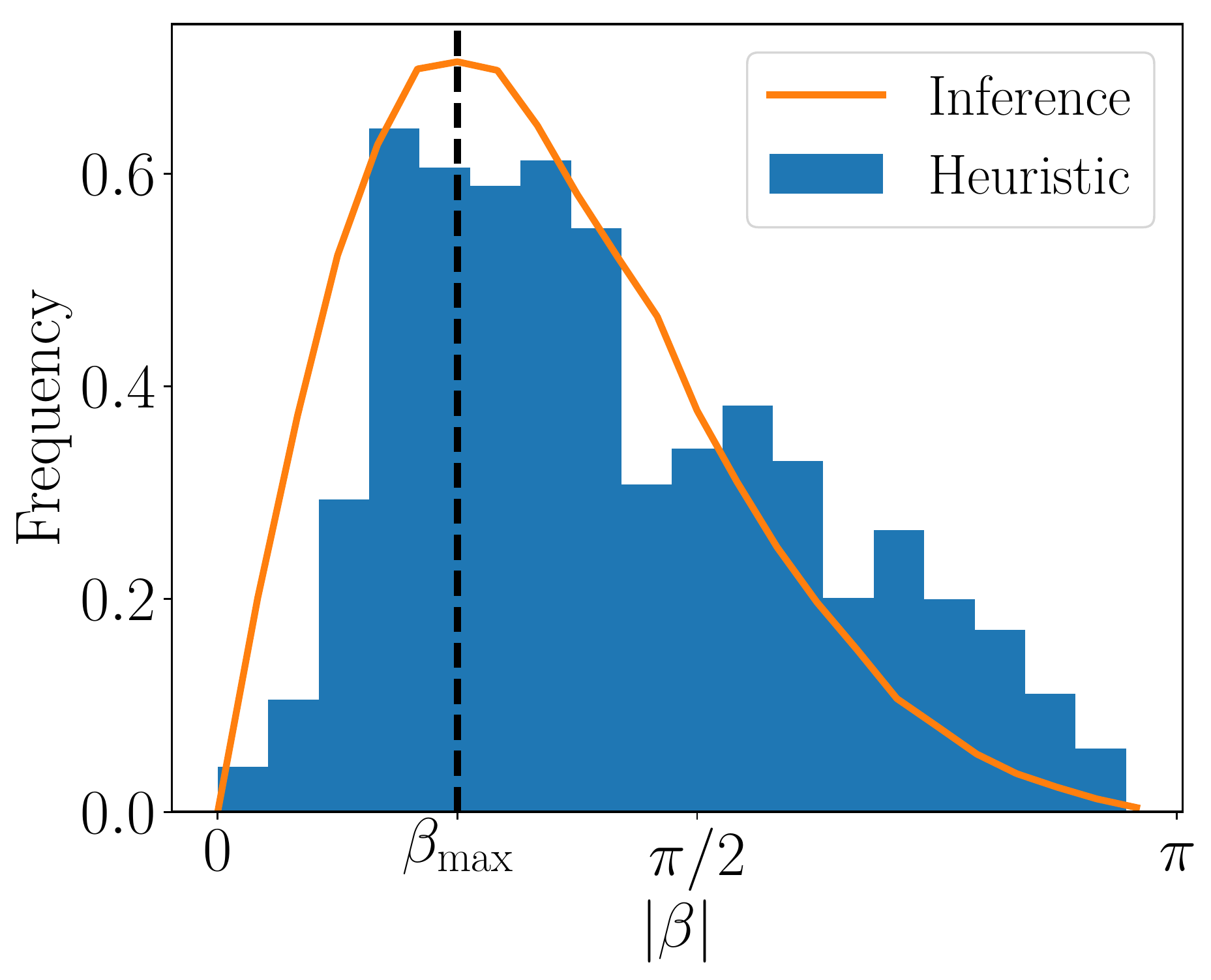}
    \caption{Comparison of the two tumble angle distributions $P(|\beta|)$
     measured by the heuristic tumble recognizer (blue bars) and determined from the stochastic process for the orientation angle using the inferred parameters $D_T$ and $r $ from table \ref{tab:Results1} (orange line). The distribution determined from theory has a maximum at $\beta_{\mathrm{max}} = 0.78 = \ang{45}$ and the mean tumble angle is $\langle |\beta |\rangle = 1.06 = \ang{61}$.}
     \label{fig:TumbleAngleDistr}
\end{figure}

\begin{figure}[h]
  \includegraphics[width= 0.5\textwidth]{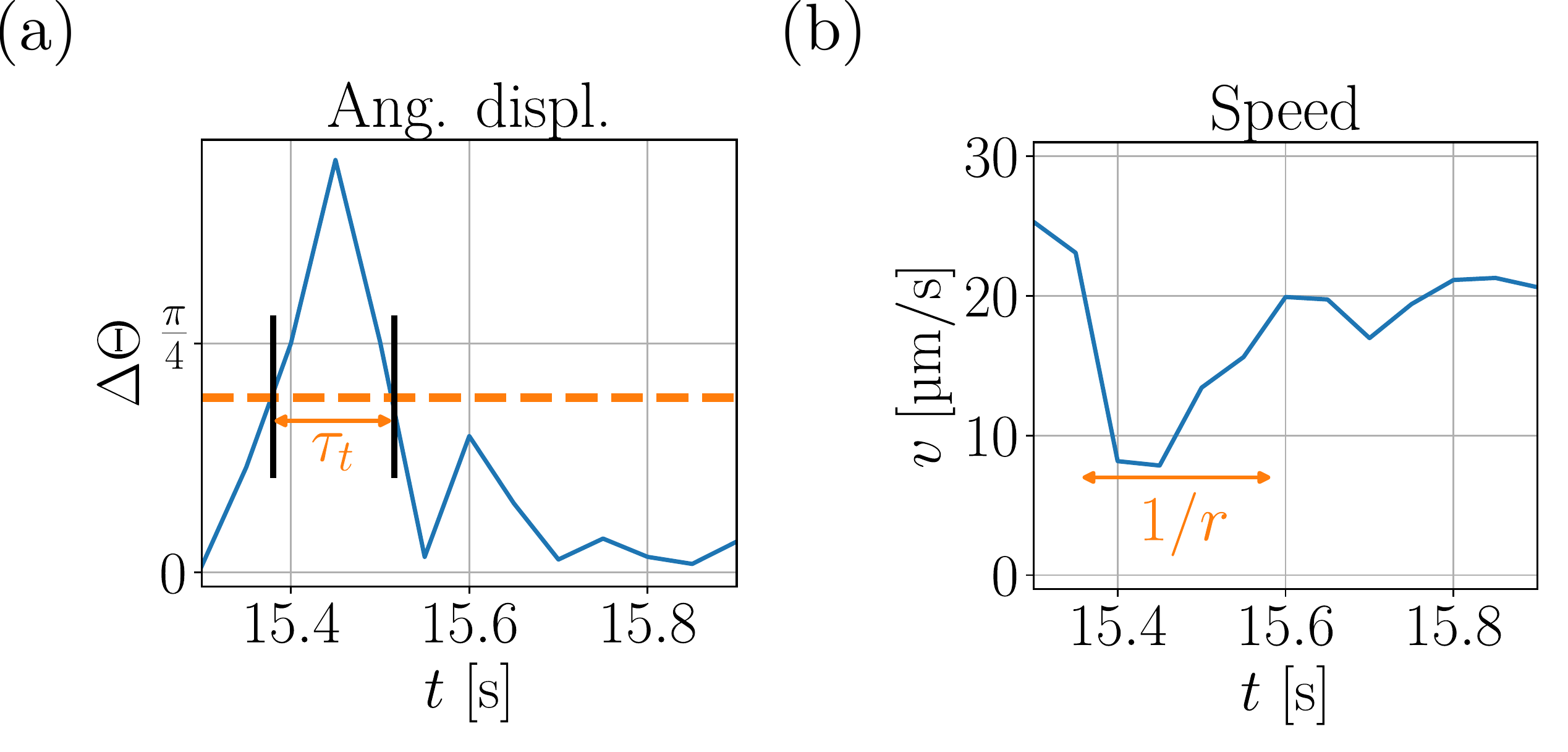}
  \caption{(a) Usually, the tumble time $\tau_t$ is defined as the period where the angular displacement per time step exceeds an a-priori threshold value.
   (b) In our method the tumble time is the inverse speed relaxation rate $r^{-1}$.}
  \label{fig:rationalChoice}
\end{figure} 

Compared to literature we define the tumble time differently by setting $\tau_t = r^{-1}$. Usually, one employs a tumble recognizer and identifies the tumble state when the angular displacement (per time step) exceeds a threshold value \cite{berg1972chemotaxis,turner2000real,turner2016visualizing,saragosti2011directional}. 
The duration of this period is then the tumble time [see also Fig.\ \ref{fig:rationalChoice}(a)], for which values of  $\tau_t = \SI{0.12}{s}$ and $\SI{0.14}{s}$ were measured using different thresholds \cite{saragosti2011directional,berg1972chemotaxis}.
However, this procedure underestimates the duration of a tumble event, which starts when a flagellum leaves the bundle and ends when it returns to the bundle. At the beginning and end of this period the angular displacement (per time step) can of course be below the given threshold value. Indeed, Ref.\ \cite{turner2000real} showed that the duration of a tumble event obtained from 
visualizing the flagellar dynamics during tumbling is significantly larger than the time determined by tumble recognizers.

In contrast to tumble recognizers, our method defines the tumble time as the inverse relaxation rate $\tau_t = r^{-1}$. This is a more rational quantification of the tumble time without the need of an a-priori threshold value. Tumbles are initiated when the speed jumps below the swimming speed and they end when the speed has relaxed back to the swimming speed. We argue that the higher value $\tau_t = \SI{0.23}{s}$ obtained by our method describes the tumble process more precisely.

\subsection{Chemotaxis}
\label{Conditioning}

Next, we apply our method to experimental data of \ec recorded in a constant gradient of a
chemoattractant concentration. Conditioning the analysis on the swimming direction, we are able to determine how the swimming parameters depend on the orientation or swimming angle $\theta$.
Thus, we divide the experimental data into eight subsets or sectors
each spanning a range of orientation angles centered at $\theta_n= 2\pi n / 8$ for $n= \{0,1,...,7 \}$. Here, $\theta = 0,2\pi$ means swimming up the gradient and $\theta = \pi$ against the gradient. In practice, instead of dividing the data for the orientation angle into 8 disjunct sectors, we use smooth weighting based on Gaussian kernels as in Ref. \cite{pohl2017inferring} (for further details see appendix \ref{GaussianKernelWeighting}).

Figure \ref{fig:ChemotaxisSwimmingDirection} shows the results from applying our inference method to the moments of speed and to the distribution of angular displacements.
Graph (a) plots the tumble bias $\lambda/r$, the ratio of tumble time to run time, versus orientation angle. It is lowered when swimming up the gradient ($\theta = 0, 2\pi$) and increased when swimming down the gradient ($\theta = \pi$).
This confirms the classical chemotaxis strategy. The curves from angle inference (orange) and speed inference (blue) show good agreement. Again, we recognize that both inference strategies give coherent results, even though they are performed independently
from each other. 
In Fig.\ \ref{fig:ChemotaxisSwimmingDirection}(b) the rotational diffusion coefficient $D_T$ 
during tumbling also depends on the swimming direction.
It is lowered when swimming up the gradient and increased when swimming down the gradient.
This suggests angular persistence or a reduced mean tumble angle, when swimming in a favorable direction,
as a chemotaxis strategy.
It was already reported in Refs. \cite{saragosti2011directional,pohl2017inferring}. We will comment more on this strategy in the following.

\begin{figure}
  \includegraphics[width=0.49\textwidth]{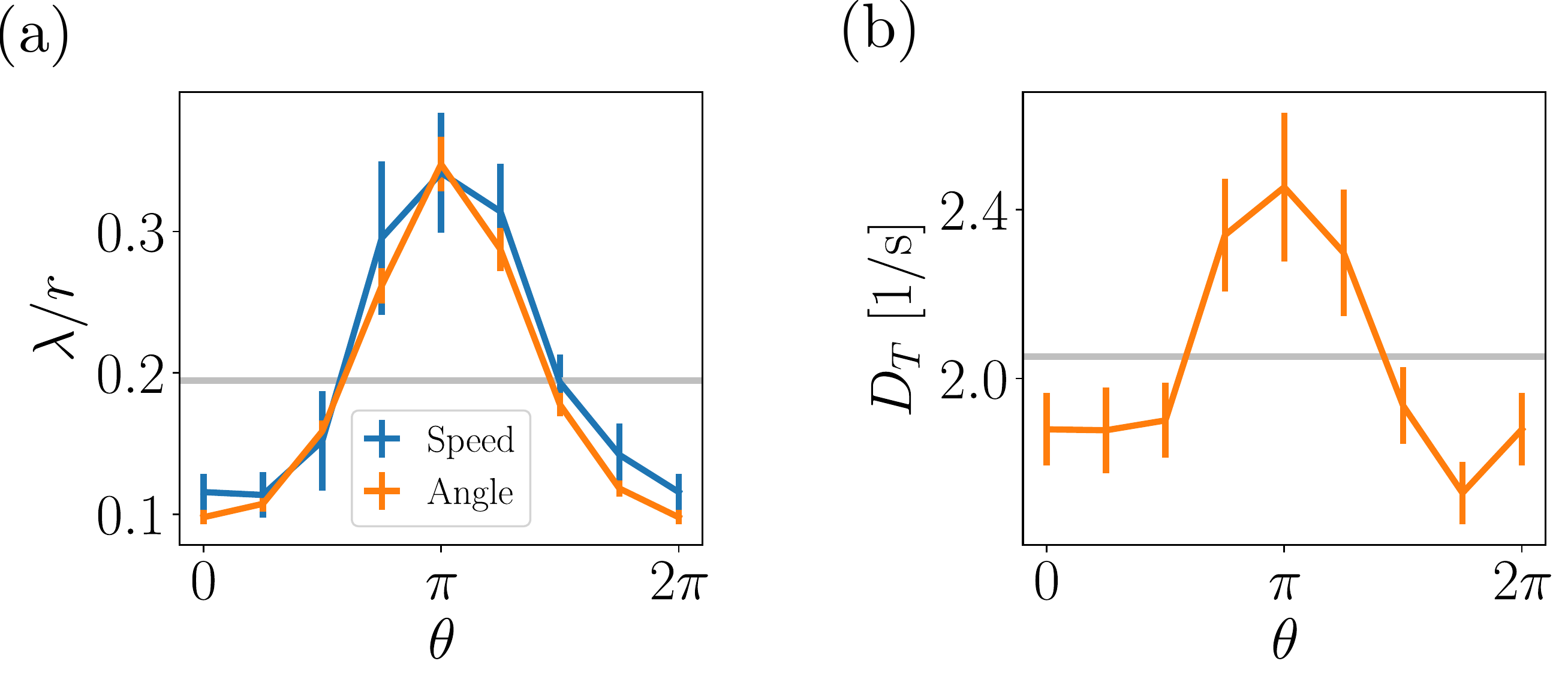}
  \caption{(a) Tumble bias $\lambda / r$ conditioned on the swimming angle $\theta$ and determined either
    by angle inference (orange) or speed inference (blue) for \ec in a linear gradient of chemoattractant ($\alpha$-methyl-aspartate). (b) Rotational diffusion coefficient $D_T$ during tumbling conditioned on the orientation angle $\theta$. The bacterium swims up the gradient for $\theta = 0, 2\pi$ and down the gradient for $\theta = \pi$.}
  \label{fig:ChemotaxisSwimmingDirection}
\end{figure}

Adding the speed autocorrelation function to the parameter inference, we investigate whether tumble rate $\lambda$ and tumble time $r^{-1}$ are separately modulated during chemotaxis.
Figure\ \ref{fig:ChemotaxisSwimmingDirectionSpeed4Param} shows the results for the speed parameters 
$\lambda, r, v_0, \sigma$. Indeed, we recover the classical chemotaxis strategy in plot (a) with a strong reduction of the tumble rate when swimming up the chemical gradient.
The tumble rate for $\theta = 0$ is less than half of the tumble rate for $\theta = \pi$.  The same trend occurs for the tumble time $r^{-1}$, which increases when swimming down the gradient.
This bias in tumble time together with the same trend for the diffusion coefficient $D_T$ found above confirms a bias in the mean tumble angle $\langle \beta \rangle$. It is enhanced when swimming in an unfavorable direction, which confirms the alternative chemotaxis strategy identified in Refs.\ \cite{saragosti2011directional,pohl2017inferring}. No significant modulations are visible for the swimming speed $v_0$ plotted in (c). So there is no chemokinesis. The same applies to the jump height $\eta$, which is shown in Fig.\ \ref{fig:JumpSizesCondition} of appendix \ref{JumpSizesCondition}. In the last plot (d) we identify a novel bias in speed fluctuations. The swimming speed is significantly more volatile when swimming up a chemical
gradient compared to swimming against it. To the best of our knowledge, this has not been reported yet. 
\begin{figure}
 \includegraphics[width=0.49\textwidth]{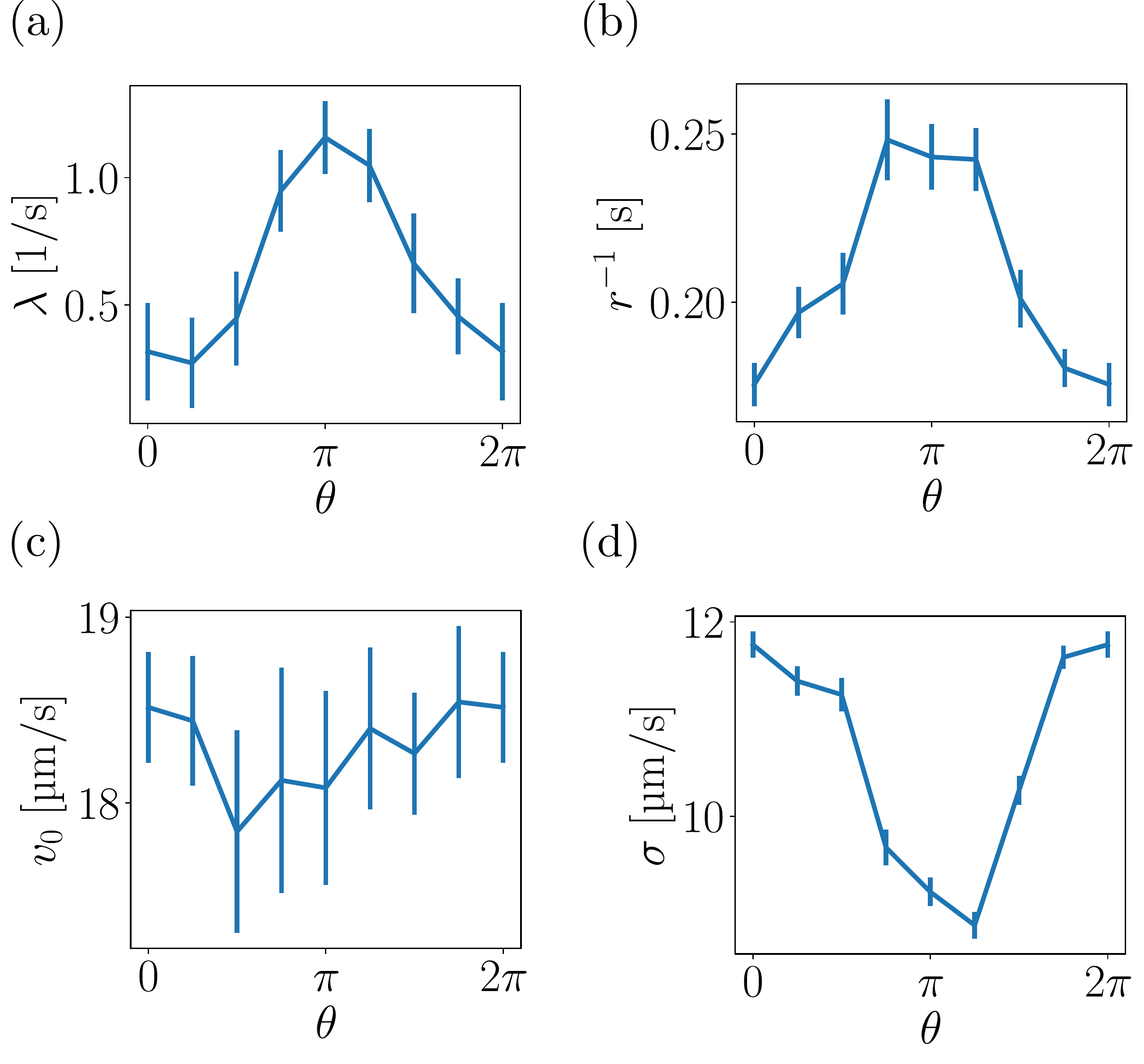}
  \caption{Inferred 
  parameters of the speed process conditioned on the swimming angle $\theta$ 
  and inferred from the same experiment as in Fig.\ \ref{fig:ChemotaxisSwimmingDirection}. We recover the bias of tumble rate in (a),
find a bias in tumble time $r^{-1}$ in (b), no chemokinesis in (c), and a novel fluctuation bias in (d). The bacterium swims up the gradient for $\theta = 0, 2\pi$ and down the gradient for $\theta = \pi$.}
  \label{fig:ChemotaxisSwimmingDirectionSpeed4Param}
\end{figure}

\section{Conclusions and Outlook}
\label{sec.concl}

In this article, we provide a detailed stochastic description of the swimming motion of an \ec bacterium in two dimensions, where we also resolve tumble events in time. We set up an overdamped Langevin equation for the speed dynamics, which contains three terms associated with drift, diffusion, and jumps that initiate a tumble event. A second Langevin equation for the angular dynamics describes rotational diffusion of the orientation angle, where the diffusion coefficient alternates between its thermal value during run phases and an enhanced value during tumbling. The transition between both phases is described by 
a telegraph process. An analysis of experimental data verifies our description a-posteriori: distribution and autocorrelation functions 
for both speed and orientation angle agree with theoretical predictions from our model and with numerically determined
functions using the inferred swimming parameters.

We considerably extent earlier work \cite{pohl2017inferring} by resolving tumble events in time and by incorporating a stochastic process for the speed dynamics. Based on moments as well as distribution and autocorrelation functions, we provide a robust methodology for inferring the full set of swimming parameters that characterize the run-and-tumble motion. The inferred swimming parameters are the tumble rate $\lambda$, the tumble time $r^{-1}$, the swimming speed $v_0$, the strength of speed fluctuations $\sigma$, the jump height $\eta$, the thermal value for the rotational diffusion coefficient $D_0$, and the enhanced coefficient during tumbling $D_T$. Although the inference of angle and speed parameters are carried out completely independent from each other, they show good and very good agreement for the two common swimming parameters, $r$ and $\lambda$, respectively.

We validated our results by comparing the swimming parameters to the results of a heuristic tumble recognizer and obtained good agreement. However, our approach of inferring parameters has three advantages. First, it does not need to set a-priori threshold parameters for speed and angular displacement. Second, it is able to infer the strength $\sigma$ of speed fluctuations and the
thermal rotational diffusion coefficient $D_0$. Third, it provides a more rational and precise choice for the tumble time that encompasses the whole tumble event instead of just the part which is determined by threshold parameters.

The inference method allows to condition the swimming parameters on a specific situation and monitor how they change with the situation by dividing the full data set into subsets. In particular, while conditioning on the swimming direction, we are able to confirm the classical chemotaxis strategy, which modulates the tumble rate $\lambda$ when changing the swimming direction relative to the chemical gradient. We also confirm the recently discovered modulation of the mean tumble angle (angle bias) \cite{saragosti2011directional}. Resolving the tumble event in time, we realize that this angle bias is due to modulations of both the tumble time and the enhanced rotational diffusivity during tumbling. This has not been reported so far. As the tumble rate we expect the tumble time to be determined by the internal chemotaxis machinery of \ec, which monitors the changing chemoattractant concentration during swimming. The higher rotational diffusivity during a tumble phase, which follows swimming against the gradient, may be caused by more flagella leaving the flagellar bundle, as argued in \cite{Saragosti2012}. Finally and also not reported so far, we show that speed fluctuations are larger by 30\% when \ec swims up the chemical gradient. 

Our method of conditioning can be applied to other quantities, for example, the concentration $c$ of the chemoattractant. In particular, the tumble rate of a bacterium, which is adapted to a chemoattractant, should not depend on the concentration $c$ \cite{macnab1972gradient}. In an earlier analysis of experiments we already verified this for \ec and \textit{Pseudomonas putida} \cite{pohl2016chemotaxis}. Other possible conditions explore the biological variability in properties such as the swimming speed $v_0$ of a bacterium or its size. 

In the following we mention some further directions, where our method of inference can be applied or needs to be extented. Recent experimental techniques allow to record tracks of length of the order of $\SI{100}{s}$ \cite{taute2015high,dufour2016direct}. Such long tracks provide enough data to apply our method to a single track and thereby measure swimming parameters for individual bacteria. This can then reveal and quantify heterogeneities in a bacterial population.

To apply the method of inference to other bacterial swimming mechansim, the Langevin equations (\ref{eq:vt}) and (\ref{eq:theta}) need to be modfied. For example, run-reverse bacteria such as the soil bacterium \textit{Pseudomonas putida}, possess a tumble angle distribution with a sharp peak centered around $\pi$ \cite{theves2013bacterial}. The marine bacteria Vibrio alginolyticus has a bimodal distribution of tumble angles with two maxima as measured in Ref.\ \cite{xie2011bacterial}. In both cases, rotational diffusion with an enhanced diffusivity cannot reproduce such distributions. A possibility to address these cases is to extend the approach of Ref.\ \cite{pohl2017inferring}. There, instantaneous tumbling was modeled by a shot noise process with a delta-peaked angular turning rate and tumble angles drawn from an 
appropriate distribution. Broadening the delta function to a Gaussian function with the tumble time $\tau_t$ as standard deviation, one can again resolve the tumble event in time. Furthermore, an elaborate model of the speed dynamics for \textit{Pseudomonas putida} should include the alternating swimming speeds reported in Ref.\ \cite{theves2013bacterial}, which belong to different swimming modes \cite{hintsche2017polar}.

Once such models are established, the inference method provides a rational way of analyzing experimental data in order to determine the relevant swimming parameters and to understand important processes such as chemotaxis by conditioning the available data on subsets. Thus, in this article we have introduce a powerful methodology for analyzing properties of bacterial populations, which can handle large amounts of experimental data.

\section*{ACKNOWLEDGMENTS} Fruitful discussions with J.-T. Kuhr, J. Blaschke, M. Hintsche and A. Kulik are acknowledged. This work was supported by the German Research Foundation (DFG) within the research training group GRK 1558.\\

\appendix
\section{Chemotaxis chamber}
\label{AppendixIbidiChamber}

Figure \ref{fig:IbidiChamber} presents a layout of the chemotaxis device used to quantify the chemotactic response of {\ec}.

\begin{figure}[h]
  \includegraphics[width=0.48\textwidth]{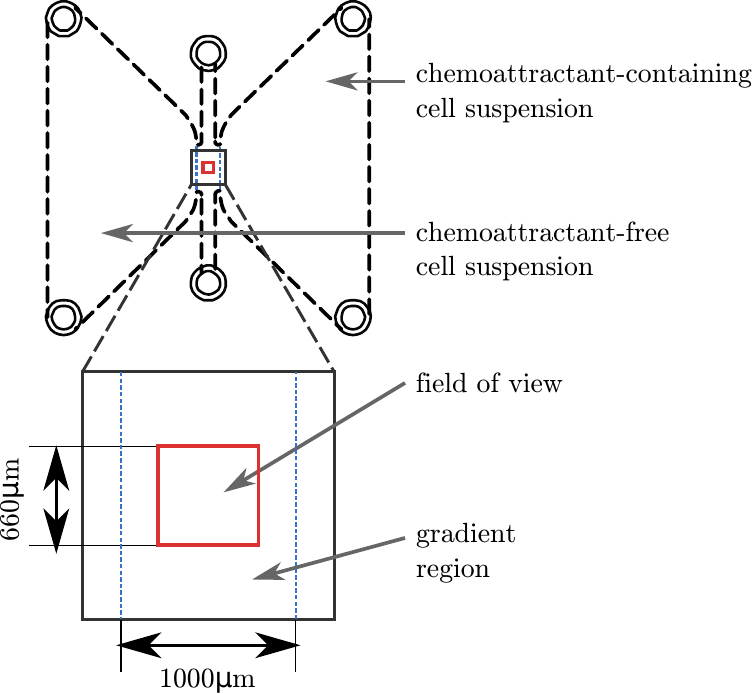}
   \caption{Layout of the chemotaxis device. The chemotaxis chamber consists of two large reservoirs connected to a central observation area. In this study, both right and left reservoirs were filled with bacterial cell suspension. The chemoattractant $\alpha$-methyl-aspartate was added to the right hand side reservoir. A linear, stable chemoattractant concentration profile was established across the central gradient region marked in blue. The bacteria were observed by video microscopy in the field of view marked in red. Figure was adapted from Ref.~\cite{pohl2017inferring}. 
   }
  \label{fig:IbidiChamber}
 \end{figure}
 
\section{Derivation of moments, autocorrelation and distribution functions}
\label{AppendixMoments}

We present detailed derivations of stochastic properties of the two Langevin equations (\ref{eq:vt}) and (\ref{eq:theta}), which we mention in the main text. First, we derive expressions for the
moments and the autocorrelation function of the speed process. Second, we present the probability distribution function (pdf), the moments, and the approximation for the directional autocorrelation function 
of the angle process.

\subsection{Speed}
\label{AppendixMomentsSpeed}

In order to perform the derivations, we rewrite Langevin equation (\ref{eq:vt}) as a stochastic differential equation (SDE) using mathematical notation:
\begin{equation}
\label{eq:vtSDE}
    dv_t = r(v_0-v_t)dt + \sigma dW_t - \eta v_t dN^{\lambda}_t ~.
\end{equation}
Here, we define the Poisson Process where $dN^{\lambda}_t =1$ occurs with probability $\lambda dt$
for each time step indicating the start of a tumble and $dN^{\lambda}_t = 0$ otherwise.
Moreover, we introduce the Wiener process $dW_t$. Integrating Eq.\ (\ref{eq:vtSDE}) and splitting the Poisson process into a deterministic part and a fluctuating part $dN_t^{\lambda} = \lambda dt + d\tilde{N}^{\lambda}_t$ \cite{cinlar2013introduction} yields

\begin{equation}
    v_t = \int_0^t r(v_0 - v_s) ds + \int_0^t\sigma dW_s - \int_0^t \eta v_s \lambda ds - \int_0^t \eta v_s d\tilde{N}^{\lambda}_t
\end{equation}
Note that the second and fourth term on the RHS are martingales\ \cite{cinlar2013introduction}.
Thus, their expectation values vanish. We will use this property when calculating the moments and autocorrelation function of the speed variable.
Taking the expectation value $\langle \ldots \rangle$ on both sides, we obtain the first moment:
\begin{equation}
   m_1 = \langle v_t \rangle = rv_0t   -     \int_0^t \left(r+\eta\lambda\right)\langle v_s\rangle ds~ .
\end{equation}
To ease the notation, we dropped the superscript $V$ from the main text. Taking the time
derivative on both sides, we obtain a non-homogeneous ordinary differential equation (ODE):
\begin{equation} 
    \frac{dm_1}{dt}  = rv_0 - (r+\eta\lambda) m_1
\end{equation}
Its full solution with initial value $C$ at time $t_0$ reads
\begin{equation}
\label{eq:speedMoment1Full}
    m_1(t) =  \frac{v_0}{1+\eta\frac{\lambda}{r}} + e^{-(r+\eta\lambda)(t-t_0)} \left( C - \frac{v_0}{1+\eta\frac{\lambda}{r}} \right)~.
\end{equation}
Taking the long-time limit $t \rightarrow \infty$, we recover the equation (\ref{eq:SpeedMoment1}) from the main text.

Next, we calculate the $n$-th moment $m_n =  \langle v^n\rangle$. Using Ito's lemma \cite{rogers1994diffusions}, we first formulate a SDE for an arbitrary function $f(v_t)$ of the speed variable:
\begin{align} 
\label{eq:ItosLemma}
\begin{aligned}
df(v_t) = & \left(f'(v_t) r(v_0-v_t) + \frac{1}{2}f''(v_t) \sigma^2 \right) dt  \\
          & + f'(v_t)\sigma dW_t  + \left[f( v_{t-} - \eta v_{t-}) - f(v_{t-})\right]dN_t^{\lambda}\, .
\end{aligned}
\end{align}
Here, $v_{t-}$ denotes the value right before a jump. Setting $f(v_t) = v_t^n$, integrating Eq.\ (\ref{eq:ItosLemma}), and taking the expectation value on both sides yields:
\begin{align} 
\label{eq:ItosLemma2}
\begin{aligned}
 \langle v_t^n \rangle =& \int_0^t \biggl( nrv_0 \langle v_s^{n-1}\rangle +(\lambda\left[(1-\eta)^n -1)\right] - nr) \langle v_s^{n} \rangle \biggr.\\
              &~~~~~~~+ \biggl. \frac{n(n-1)}{2}\sigma^2 \langle v_s^{n-2} \rangle \biggr) ds\, ,
\end{aligned}
\end{align}
where we again extracted the deterministic part of the Poisson process and all martingales dropped out. Taking the time derivative on both sides, we obtain an ODE, which also contains the lower-order moments $m_{n-1}$ and $m_{n-2}$:
\begin{align} 
\label{eq:ItosLemmaODE}
\begin{aligned}
\frac{d m_n}{dt}  =&~ nrv_0~ m_{n-1} +\biggr(\lambda\left[(1-\eta)^n -1\right] - nr\biggr) ~m_{n}\\
              &~~~~~~~+  \frac{n(n-1)}{2}\sigma^2 ~m_{n-2}.
\end{aligned}
\end{align}
The solution of this ODE in the long-time limit $t \rightarrow \infty$, where $d m_n / dt = 0$,
yields Eq.\ (\ref{eq:SpeedMomentN}) in the main text,
\begin{equation}
    m_n = \frac{v_0 ~m_{n-1} ~+~ \frac{1}{2} \left(n-1\right) \frac{\sigma^2}{r} ~m_{n-2}  } 
       {1 + \frac{\lambda}{n r} - \frac{\lambda}{nr} \left( 1-\eta \right) ^n } ~.
\end{equation}

Finally, we calculate the speed autocorrelation function $g(s,t) = \langle (v_{s} - m_1) (v_t - m_1 )\rangle$ of Eq.\ (\ref{eq:vtSDE}). We define the probability distributions for the speed process $P(v')$ and the conditional probability $P(v,t|v',s)$ of having $v$ at time $t$ given that we have $v'$ at time $s$ and obtain 
\begin{align} 
\label{eq:SpeedAutocorrDerivation}
\begin{aligned}
g(s,t)     &= \int\int \left[ (v-m_1)(v'-m_1) P(v,t|v',s)P(v') \right]dvdv'\\
           & = \int\left[   \langle v(t)-m_1|[v',s]\rangle           (v'-m_1)    P(v')\right] dv'   \\
           & = \int \left[ \left((v' - m_1)e^{-(r+\eta\lambda)|t-s|} \right) (v'-m_1) P(v')\right]dv' \\
           &= \Delta v^2 ~e^{-(r+\eta\lambda)|t-s|}~,
\end{aligned}
\end{align}
where we have have used Eq.\ (\ref{eq:speedMoment1Full}) with $C=v'$ in the second last step. We recover Eq.\ (\ref{eq:SpeedAutocorr}) after setting $s = t+\tau$. Identifying the relaxation rate $\alpha_V$, we can write the following formulas for $\lambda$ and $r$:
\begin{align}
    \lambda =&  \frac{\alpha_V}{1+\eta\lambda/r}~,\\
    r       =& \frac{\alpha_V}{\eta+(\lambda/r)^{-1}}~.
\end{align}

\subsection{Angle}
\label{AppendixAngle}

We rewrite the Langevin equation (\ref{eq:theta}) from the main text as a SDE using mathematical notation:
\begin{equation}
d\Theta_t = \sqrt{2D_t}dW_t
\end{equation}
The SDE contains two stochastic processes: the telegraph process $D_t$, where we drop here the subscript $\rm{rot}$ 
used in the main text, and the white noise process $dW_t$. These two processes are stochastically independent of each other. Thus, the moments for the angular displacement during time step $\Delta t$ factorize into contributions from each process,
\begin{equation}
\label{eq:AngleMomentsFactorize}
    \left\langle |\Delta\Theta|^n \right\rangle = \left\langle [2D_t]^{\frac{n}{2}}\right\rangle\left\langle |\Delta W_t|^n\right\rangle~.    
\end{equation}
The probability distribution function (pdf) $p(\Delta W_t)$ and the absolute moments of the white noise increments $\Delta W_t$ during time step $\Delta t$ are given by
\begin{align} 
    p(\Delta W_t) &= \mathcal{N}(0,\sqrt{\Delta t})~,\label{eq:BrownianPDF}\\
     \langle |\Delta W(t)|^n\rangle &= (\Delta t)^{\frac{n}{2}} (n-1)!!\begin{cases} \sqrt{\frac{2}{\pi}} ~~~&\textrm{if $n$ is odd}\\ 1 ~~~&\textrm{if $n$ is even} \end{cases}~,\label{eq:BrownianAbsoluteMoments}
\end{align}
where $\mathcal{N}(0,\sigma)$ denotes the normal distribution with zero mean and standard deviation $\sigma$ and $n!!$ denotes the double factorial.

For the telegraph process $D_t$ with states $D_0$ and $D_T$, the two probabilities for being in one of the states at time $t$ obey the following master equations:
\begin{align}
\partial_t P(D_0,t|C,t_0) &= -\lambda P(D_0,t|C,t_0) + r P(D_T,t|C,t_0) \, , 
\nonumber \\
\partial_t P(D_T,t|C,t_0) &=  ~~~\lambda P(D_0,t|C,t_0) - r P(D_T,t|C,t_0) \, .
\nonumber
\end{align}
Here,  $\lambda$ is the transition rate from $D_0$ to $D_T$ and $r$ the transition rate for the reverse process. The variable $C$ indicates the initial condition at time $t_0$. We first state the pdf $p(D)$ in the long-time limit $t\rightarrow \infty$ as well as the auto-correlation function $\langle D_tD_s \rangle$ from literature \cite{lindner2009brief}:
\begin{align}
p(D_0) &=  \frac{r}{\lambda+r} ~,\label{eq:TwoStatePDF1}\\
p(D_T) &=  \frac{\lambda}{\lambda+r}~,\label{eq:TwoStatePDF2} \\
\langle D_tD_s \rangle &= \langle D \rangle ^2 + \Delta ^2 D~ e^{-(\lambda+r)|t-s|}~.
\label{eq:TwoStateAutocorr}
\end{align}
In the last equation we have introduced the mean $\langle D \rangle$ and the variance $\Delta ^2 D$ in the long time limit. They are given by
\begin{align}
\langle D \rangle &= \frac{D_0 r + D_T\lambda}{\lambda+r}~,\\
\Delta ^2 D &= \frac{(D_0 -D_T)^2 \lambda r}{(\lambda+r)^2}~.
\end{align}
The mean value of $D_t$ for any time $t$ with initial condition $C$ at time $t_0$ is given by 
\begin{equation}
    \langle D_t \rangle = \langle D \rangle  + e^{-(\lambda+r)(t-t_0)} \left( C - \langle D \rangle \right)~.
\end{equation}
We can use the pdf $p(D)$ to calculate the first factor on the RHS of Eq.\ (\ref{eq:AngleMomentsFactorize}) in the long time limit,
\begin{equation}
    \left\langle [2D_t]^{\frac{n}{2}}\right\rangle = \frac{(2D_0)^{\frac{n}{2}}}{1+\lambda /r} + \frac{(2D_T)^{\frac{n}{2}}}{1+r/\lambda}~.
\end{equation}
Inserting this expression and Eq.\ (\ref{eq:BrownianAbsoluteMoments})
in Eq.\ (\ref{eq:AngleMomentsFactorize}) leads to Eq.\ (\ref{eq:TurnrateMomentN}) stated in the main text.

The pdf of the absolute angular displacement $p(|\Delta\Theta|)$ can be calculated straightforwardly. Using the independence of the two stochastic processes and combining Eqs.\ (\ref{eq:BrownianPDF}), (\ref{eq:TwoStatePDF1}), and (\ref{eq:TwoStatePDF2}), we obtain
\begin{equation}
        p(\left|\Delta\Theta\right|) = \frac{r}{\lambda + r} \mathcal{N}(0,\, \sqrt{2D_0 \Delta t}) + \frac{\lambda}{\lambda + r} \mathcal{N}(0,\,\sqrt{2D_T \Delta t})~.
\end{equation}
This agrees with Eq.\ (\ref{eq:pdfTurnrate}) from the main text.

Finally, we calculate the directional autocorrelation function $g(\tau) = \langle \bold{e}(\tau) \cdot \bold{e}(0)\rangle = \langle \cos\left(\Theta(\tau) -\Theta(0)\right)\rangle $.
 Integrating Eq.\ (\ref{eq:theta}) and using the real part $\Re$ of the Euler identity $e ^{ix}= \cos(x) + i\sin(x)$ yields:
\begin{align}
   g(\tau) 
   &= \Re ~\left\langle e^{i\int_{0}^{\tau} \sqrt{2D_s} dW_s }\right\rangle
\end{align}
The term in the real part operator can be interpreted as the characteristic function of the random variable $X(\tau) = \int_{0}^{\tau} \sqrt{2D_s}dW_s$ for wavenumber $k=1$.
Using the moment representation of the characteristic function, we obtain
\begin{equation}
\label{eq:SumAngleAutocorr}
    g(\tau) = \Re\sum_{n=0}^\infty \frac{i^n}{n!} m_n(\tau) \, ,
\end{equation} 
where we have defined the moments $m_n = \langle X^n \rangle$. For symmetry reasons, the odd moments vanish,
\begin{equation}
    m_{2n+1} = 0 \, ,
\end{equation}
and the real part operator can be skipped. First, we calculate $m_2$, where we use again the independence of the two stochastic processes $dW_t$ and $D_t$ in the second line,
\begin{align}
    m_2(\tau) &= \langle \int_{0}^{\tau} \sqrt{2D_{s_1}}dW_{s_1}\int_{0}^{\tau} \sqrt{2D_{s_2}}dW_{s_2}\rangle \nonumber \\
          &=  \int_{0}^{\tau}  \int_{0}^{\tau} \langle\sqrt{2D_{s_1}}\sqrt{2D_{s_2}}\rangle\langle dW_{s_1}dW_{s_2}\rangle \nonumber \\
          &=  \int_{0}^{\tau}  \int_{0}^{\tau} \langle\sqrt{2D_{s_1}}\sqrt{2D_{s_2}}\rangle \delta(s_1-s_2)d{s_1}d{s_2} \nonumber \\
          &=  2\int_{0}^{\tau} \langle D\rangle d{s_1} \nonumber \\          
          &= 2 \langle D \rangle \tau
\end{align}
Next, we calculate the fourth moment $m_4$, where we use the correlation function of Eq.\ (\ref{eq:TwoStateAutocorr}) in the fourth line:
\begin{widetext}
\begin{align}
     m_4(\tau) &=  \int_{0}^{\tau}  \int_{0}^{\tau}  \int_{0}^{\tau}  \int_{0}^{\tau}  \langle\sqrt{2D_{s_1}}\sqrt{2D_{s_2}}\sqrt{2D_{s_3}}\sqrt{2D_{s_4}}\rangle
                \langle dW_{s_1}dW_{s_2}dW_{s_3}dW_{s_4}\rangle   \nonumber \\
            &=  \int_{0}^{\tau}   \int_{0}^{\tau}  \int_{0}^{\tau}  \int_{0}^{\tau}  \langle\sqrt{2D_{s_1}}\sqrt{2D_{s_2}}\sqrt{2D_{s_3}}\sqrt{2D_{s_4}}\rangle
                3\delta(s_1-s_2)\delta(s_3-s_4) ds_1ds_2ds_3ds_4    \nonumber \\ 
            &=  12 \int_{0}^{\tau}   \int_{0}^{\tau}  \langle D_{s_1}D_{s_2}\rangle ds_1ds_2  \nonumber \\ 
            &= 12 \int_0^t  \int_0^t  \langle D \rangle ^2 + \frac{(D_0 -D_T)^2 r\lambda }{(\lambda+r)^2} e^{-(\lambda+r)| s_1-s_2|}  ds_1ds_2 \nonumber \\
            &= 12\left(\langle D \rangle ^2 \tau ^2  + I \right) \nonumber \\
            &=  12 \left(\langle D \rangle ^2 \tau ^2 + 2 \frac{\Delta ^2 D}{\lambda+r} \left[ \tau + \frac{ e^{-(\lambda+r)(\tau-0)}}{\lambda+r}  - \frac{1}{\lambda+r} \right] \right)
\end{align}
\end{widetext}
Replacing in the double integral $\int_0^\tau \ldots ds_1$ by $2\int_0^{s_2} \ldots ds_1$,  the integral $I$
is calculated as follows:
\begin{align}
    I &= 2 \int_{0}^{\tau}   \int_{0}^{s_2} \Delta ^2 D e^{-(\lambda+r)(s_2-s_1)}  ds_1ds_2 \nonumber \\
      &= 2 \int_{0}^{\tau}  \left[ \frac{\Delta ^2 D}{\lambda+r}  e^{-(\lambda+r)( s_2-s_1)}  \right|_{s_1=0}^{s_1=s_2}ds_2 \nonumber \\
      &= 2 \frac{\Delta ^2 D}{\lambda+r}\int_{0}^{\tau}  1 -  e^{-(\lambda+r)s_2}ds_2 \nonumber \\ 
      &= 2 \frac{\Delta ^2 D}{\lambda+r} \left[ s_2 + \frac{ e^{-(\lambda+r)s_2}}{\lambda+r}\right|_{s_2=0}^{s_2=\tau} \nonumber \\
      &= 2 \frac{\Delta ^2 D}{\lambda+r} \left[ \tau + \frac{ e^{-(\lambda+r)\tau}}{\lambda+r}  - \frac{1}{\lambda+r} \right]
\end{align}
Truncating the sum of Eq.\ (\ref{eq:SumAngleAutocorr}) for $n>4$ , we finally obtain:
\begin{equation}
\begin{aligned}
\label{eq:FinalApproximation}
    g_\Theta(\tau) = &1 -\langle D \rangle \tau + \langle D \rangle ^2 \tau ^2/2 \\
                     &+ \frac{\Delta ^2 D}{\lambda+r} \left( \tau + \frac{ e^{-(\lambda+r)\tau}}{\lambda+r} - \frac{1}{\lambda+r} \right) \, .
\end{aligned}
\end{equation}
This form suggests a slope $-\langle D \rangle$ of the correlation function for times $\tau < (\lambda+r)^{-1}$, which in our case means $\tau < \SI{0.2}{s}$ and is just valid for the very initial time range of the correlation function.
From Eq.\ (\ref{eq:FinalApproximation}) we can extract another linear approximation by concentrating on the time range 
$(\lambda+r)^{-1} < \tau < {\langle D \rangle}^{-1} $.
It gives Eq. (\ref{eq:LinearApproxAngleAutocorr}) from the main text,
\begin{equation}
    g_\Theta(\tau) = 1 - \left( \langle D \rangle - \frac{\Delta ^2 D}{\lambda+r} \right) \tau~,
\label{eq.Aalpha}
\end{equation}
from which we obtain an expression for the relaxation rate $\alpha_{\Theta}$ measured in experiments.
It is determined by $\langle D \rangle$ and the second term in the brackets is a correction. But it is sufficient to determine separate values for $r$ and $\lambda$, when $ r / \lambda$ is known from the analysis of the pdf  $p(\left|\Delta\Theta\right|)$.
Solving the equation for $\alpha_{\Theta}$ for either $\lambda$ or $r$, we obtain the formulas
\begin{align}
    \lambda =&  \frac{\Delta ^2 D}{(1+(\lambda/r) ^{-1})(\langle D\rangle -\alpha_{\Theta})}~,\label{eq:lambdaAngle}\\
    r       =& \frac{\Delta ^2 D}{(1+\lambda/r)(\langle D\rangle -\alpha_{\Theta})} \label{eq:rAngle}~.
\end{align}

\subsection{Numerical investigations of the directional autocorrelation function}
\label{NumericalVerifications}

The directional autocorrelation function $g_{\Theta}(\tau) = \left\langle \bold{e}(\tau)\cdot \bold{e}(0)\right\rangle$ has an exponential form in experiments up to ca. 2s (see Fig.\ \ref{fig:HistogramAutocorrAngle}). Here, we validate this dependence by numerically solving Eq.\ (\ref{eq:theta}) with the inferred parameters of table \ref{tab:Results1}. The semi-logarithmic plot
in Fig.\ \ref{fig:AutocorrMatchError}(a) shows the resulting autocorrelation function (blue data points). It
is in good agreement with the exponential decay of Eq.\ (\ref{eq:AngleAutocorr}) using the relacation rate $\alpha_{\Theta}$ from Eq.\ (\ref{eq:LinearApproxAngleAutocorr}), which we derived in the previous section in Eq.\ (\ref{eq.Aalpha}). This validates our proposition for the relaxation rate.

Moreover, we can further validate the exponential fit to the experimental directional autocorrelation function using the theoretical value for the relaxation rate. After having inferred the reduced parameter set $(\lambda / r, D_0, D_T)$ as described in the main text using the pdf $p(|\Theta |)$, we determine the directional autocorrelation function by simulating the angle process with the reduced 
parameter set for different values of the parameter $\lambda$. Figure \ref{fig:AutocorrMatchError}(b) shows the mean squared error $\Sigma$ of the simulated autocorrelation function compared to the experimental function plotted versus the tumble rate $\lambda$. The best match is for a $\lambda$ very close to the value shown in table \ref{tab:Results1}, which was determined using the theoretical prediction of Eq.\ (\ref{eq:LinearApproxAngleAutocorr}) for the relaxation rate $\alpha_{\Theta}$.
\begin{figure}
    \includegraphics[width=0.23\textwidth]{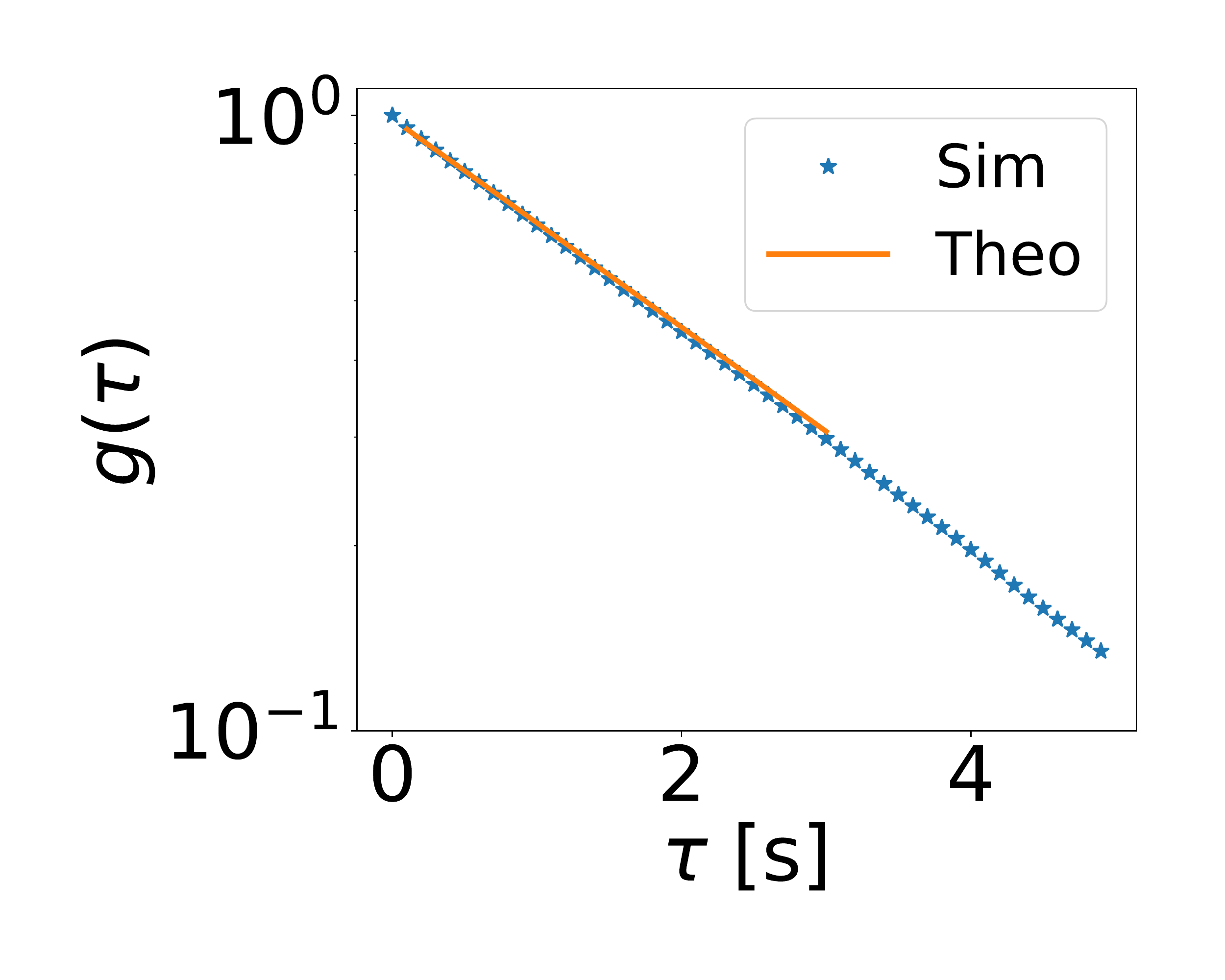}
    \includegraphics[width= 0.23\textwidth]{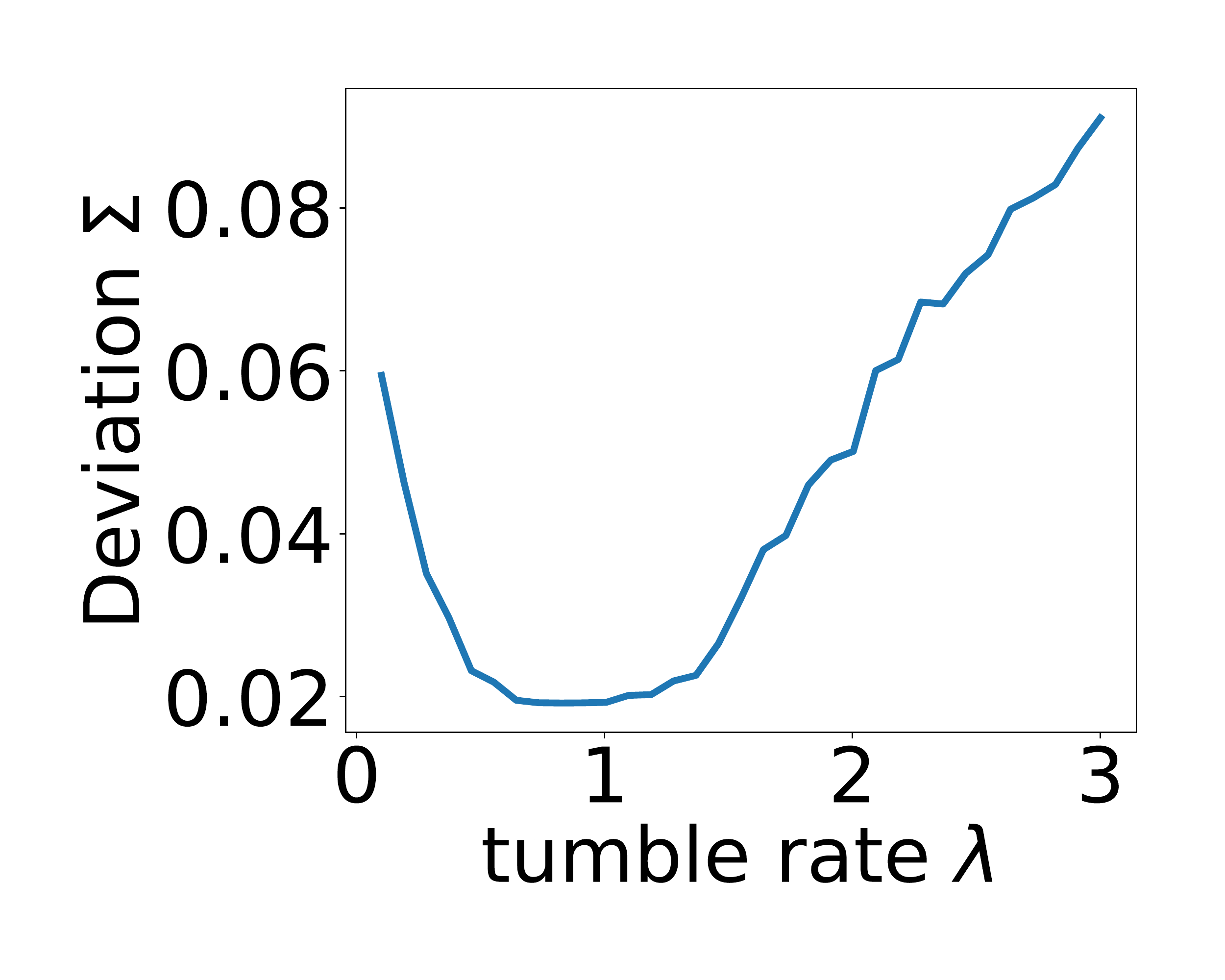}
    \caption{Left: 
    Semi-logarithmic plot of the directional autocorrelation function 
    from a numerical solution of Eq.\ (\ref{eq:theta})
    using the inferred parameters from table \ref{tab:Results1} 
    (blue data points). The orange line shows an exponential decay with the relaxation rate from 
    Eq.\ (\ref{eq:LinearApproxAngleAutocorr}). Right: Mean squared deviation between the simulated
    directional autocorrelation function $g_{\Theta}(\tau) = \left\langle \bold{e}(t+\tau)\cdot 
    \bold{e}(t)\right\rangle$ and the experimental curve for different tumble rates $\lambda$. The global minimum at $\lambda = \SI{0.81}{s^{-1}}$ verifies the use of the theoretical expression
    (\ref{eq:LinearApproxAngleAutocorr}) for the relaxation rate.
    }
    \label{fig:AutocorrMatchError}
\end{figure}

\section{The method of bootstrapping}
\label{AppendixBootstrapping}
Bootstrapping allows to derive an estimate of the standard deviation of the inferred parameters without the need of repeated experiments \cite{efron1994introduction}. 
Similar to Ref. \cite{pohl2017inferring}, we create synthetic ensembles by randomly mixing subsets
of the original data set. Let $T_0 =\{t_1,....,t_N\}$ be the set of original trajectories. 
Pulling $N$ random trajectories of this set and laying them back after each pull, one obtains a bootstrap sample $T_1 = \{\tilde{t}_1, ..., \tilde{t}_N\}$, where single trajectories can appear several times.
We create $K = 100$ of these bootstrap samples, apply our inference technique to each sample, and obtain a 
distribution of values for each swimming parameter. The error bars in the main text are the standard deviation from the mean of each swimming parameter.

\section{Smooth weighting of data}
\label{GaussianKernelWeighting}

The conditioning of section \ref{Conditioning} needs the division of the data in different sectors. Instead of a discrete division, 
we use the whole data set for each sector but weight the data by a Gaussian kernel similar 
to Ref.\ \cite{pohl2017inferring}. The speed moments for $N$ experimental trajectories when 
conditioning on a specific
swimming angle $\theta$ are then calculated according to
\begin{equation}
    \langle m^V_n \rangle  =   \frac{ \sum_{i=1}^N \sum_t v_i(t) ^n  \exp\left(-\frac{[\Theta_i(t-2\Delta t)-\theta]^2}{2\Delta\theta^2}\right)} {\sum_{i=1}^N \sum_t \exp\left(-\frac{[\Theta_i(t-2\Delta t)-\theta]^2}{2\Delta\theta^2}\right)} ~,
\end{equation}
where we have introduced the width of a section, $\Delta \theta = 0.125\pi$, 
and their centers $\theta$. Note that we use the actual orientation angle $\Theta_i(t-2\Delta t)$ of the second previous time step to calculate the moments. Tumble events have a finite duration of around $2\Delta t$ and this ensures that the whole tumble is connected to the condition of the previous run. The same Gaussian kernels are applied when we calculate the histogram of angular displacements and the autocorrelation functions for speed and direction.
\section{Jump height conditioned on swimming angle $\theta$.}
\label{JumpSizesCondition}
Figure\ \ref{fig:JumpSizesCondition} shows the relevant plot. There is no systematic dependence of $\eta$ on the swimming angle.

\begin{figure}
    \includegraphics[width= 0.23\textwidth]{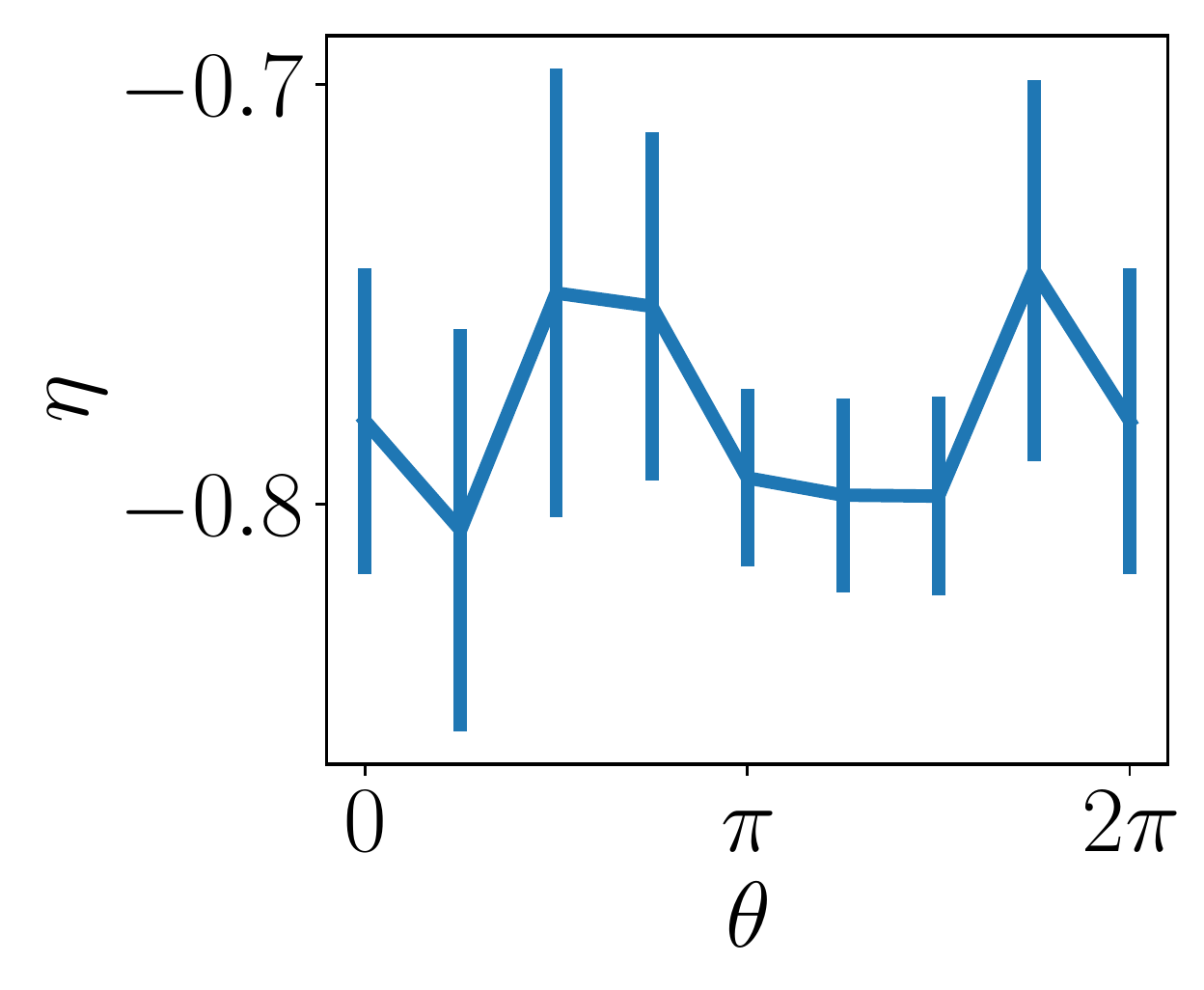}
    \caption{Jump height $\eta$ conditioned on swimming angle $\theta$.}
        \label{fig:JumpSizesCondition}
\end{figure}

\vspace*{1cm}

\bibliography{bibliography}

\end{document}